\documentclass[english,aps,prb,showpacs,twocolumn]{revtex4-1}
\usepackage[T1]{fontenc}
\usepackage[latin9]{inputenc}
\usepackage{color}
\usepackage{babel}
\usepackage{amssymb}
\usepackage{wasysym}
\usepackage{graphicx}
\usepackage[unicode=true,pdfusetitle,
 bookmarks=true,bookmarksnumbered=false,bookmarksopen=false,
 breaklinks=false,pdfborder={0 0 0},backref=false,colorlinks=true]
 {hyperref}
\hypersetup{
 citecolor=blue, linkcolor=blue, urlcolor=blue}

\makeatletter
\@ifundefined{textcolor}{}
{%
 \definecolor{BLACK}{gray}{0}
 \definecolor{WHITE}{gray}{1}
 \definecolor{RED}{rgb}{1,0,0}
 \definecolor{GREEN}{rgb}{0,1,0}
 \definecolor{BLUE}{rgb}{0,0,1}
 \definecolor{CYAN}{cmyk}{1,0,0,0}
 \definecolor{MAGENTA}{cmyk}{0,1,0,0}
 \definecolor{YELLOW}{cmyk}{0,0,1,0}
 }

\makeatother

\begin{document}

\title{Transport properties of rippled graphene}

\author{Maciej Zwierzycki}

\email{maciej.zwierzycki@ifmpan.poznan.pl}

\selectlanguage{english}%

\affiliation{Institute of Molecular Physics, Polish Academy of Sciences, Smoluchowskiego
17, 60-179 Pozna\'{n}, Poland}

\date{ver. 3}
\begin{abstract}
The exceptionally high mobility of carriers in graphene is one of
its defining characteristics, especially in view of potential applications.
Therefore it is of both practical and fundamental importance to understand
the mechanisms responsible for limiting the values of mobility. The
aim of the paper is to study theoretically one such mechanism, \emph{i.e.}
scattering on ripples. The transport properties of rippled graphene
are studied using using single-band tight-binding model. Both the
bond-length variation, corresponding to the vector potential in the
effective mass picture, and fluctuating scalar potential are included
in the formalism. The samples are modeled as self-similar surfaces
characterized by the roughness exponent with values ranging from typical
for graphene on SiO$_{2}$ to seen in suspended samples. The range
of calculated resistivities and mobilities overlaps with experiment.
The results presented here support the notion of rippling as one of
the factors limiting the mobility. 
\end{abstract}

\pacs{72.80.Vp, 61.48.Gh, 61.48.Gh}

\maketitle

\section{Introduction}

The enormous interest in graphene, a two-dimensional (2D) honeycomb
lattice of carbon atoms discovered in 2004\cite{Novoselov:sc04},
has both fundamental and practical reasons. The electronic properties
of graphene are related to the peculiar nature of its band states
in the vicinity of the Fermi level. These states possess a linear
dispersion and are described by effective Hamiltonian formally identical
to that of relativistic massless particles (Dirac fermions) albeit
with velocity 300 times lower than the speed of light.\cite{Novoselov:nat05}
This leads to a range of interesting phenomena like finite minimal
conductivity, Klein paradox (\emph{i.e.} tunneling with no attenuation)
or the anomalous Quantum Hall Effect which can be observed even at
room temperatures (see \emph{e.g.} Ref.~{[}\onlinecite{Neto:rmp08}{]}
for a review). The practical interest stems mostly from the fact that
carriers in graphene, whose number and character (electrons or holes)
can be modified by field effect, possess exceptionally high mobility.
The values range from order of $10.000-20.000\;\mathrm{cm^{2}/(V\cdot s)}$
for exfoliated graphene on SiO$_{2}$ substrate to over $100.000\;\mathrm{cm^{2}/(V\cdot s)}$
for suspended samples,\cite{Bolotin:prl2008,Du:NNano2008} making
graphene a strong candidate for use in ultra-fast electronic devices.\cite{Lin:NanoLet2009,Liao:Nature2010,Lin:Science2011,Schwierz:NatNano2010} 

The precise mechanism limiting the mobility of carriers in graphene
is still a matter of debate. The values and temperature dependence
of mobilities measured in suspended samples (in the doped regime)
seem to support the phonon scattering model.\cite{Bolotin:prl2008}
The situation is less clear in samples on substrate where little temperature
dependence is observed. An often suggested source of scattering in
the latter case are charged impurities, possibly located in the substrate.\cite{adam:ssc2009}
However, an experiment designed specifically to test this scenario
failed to produce supporting results.\cite{ponomarenko:prl2009} Another
possibility is scattering induced by ripples,\cite{katsnelson:pht08}
which has been observed in both suspended\cite{Meyer:nat07} and exfoliated
samples.\cite{Stolyarova:pnas07,ishigami:nanolet07,geringer:prl09,Cullen:prl2010}
Theoretical studies indicate that ripples can be caused by direct
interaction with the substrate,\cite{Cullen:prl2010} thermodynamical
instability inherent to 2D systems\cite{fasolino:natm07,Abdepour:prb2007},
functionalisation \emph{e.g.} by OH groups\cite{Kim:EPL2008} or doping.\cite{gazit:prb2009}
The picture which emerges is that of graphene as self-similar rippled
membrane. 

One way to characterize such structure is via height-correlation function
$\left\langle (h(\mathbf{r})-h(\mathbf{0}))^{2}\right\rangle $ which
exhibits $r^{2H}$ scaling for small values of $r$ and eventually
saturates at twice the standard deviation of height, $std(h)=\sqrt{\left\langle h^{2}\right\rangle }$
(we assume $\bar{h}=0$). The scaling exponent $H$ is related to
the fractal dimension of the rippled surface and assumes values ranging
from $0.5$ for exfoliated\cite{ishigami:nanolet07} to $0.7-1.0$
for suspended\cite{fasolino:natm07,Abdepour:prb2007,Los:prb2009}
samples. The larger values of H in the latter case indicate locally
smoother structures. The lateral size of surface features can be characterized
by the correlation length $\xi$, defined as rollover point of height-correlation
function%
\footnote{The height correlation function typically scales like $\left< (h(\mathbf{r})-h(0))^2\right>\sim 2\left<h^2\right>\left( 1 - e^{(r/\xi)^{2H}} \right)$%
}, varying between several and few tens of nanometers.\cite{Abdepour:prb2007,ishigami:nanolet07,Cullen:prl2010}
Finally, the size of out of plane fluctuations can be described by
standard deviation of height, $std(h)$, usually equal to few Angstroms.

The most direct mechanism by which rippling modifies the electronic
structure of graphene is the alteration of bond lengths (and consequently
hopping integrals). Interestingly, in the effective mass approximation
the resulting modifications to the Hamiltonian have a form of a random
gauge field,\cite{guinea:prb08a,guinea:prb08b,vozmediano:physrep2010}
whose presence may be responsible for suppression of weak localization
observed in some graphene samples.\cite{Morozov:prl06} Furthermore
the presence of ripples can lead to the fluctuations of charge density
and in particular formation of electron and hole puddles in globally
neutral samples. Gibertini \emph{et al.}\cite{gibertini:prb10} calculated
deformation potential for geometries obtained from Monte Carlo simulations\cite{fasolino:natm07,Los:prb2009}
and found it and the local charge density to vary on length scale
of 1 -- 2~$\mathrm{nm}$ with no apparent correlation to the height
variation.%
\footnote{Longer length scales were obtained by the same group\cite{gibertini:prb(r)2012}
when their formalism was applied to the experimental structure of
Ref.~\onlinecite{geringer:prl09}.%
} Similar length scales were also obtained in \emph{ab initio} calculations
of Ref.~\onlinecite{partovi:prb2011}. Kim and Castro\cite{Kim:EPL2008}
and Gazit\cite{gazit:prb(R)2009} on the other hand obtained close
correlation between the charge fluctuations and and the local curvature
of the graphene sheet. In this case the effective scalar potential
varies on the length scales comparable to the coherence length. In
all cases the amplitudes of the scalar potential variation were on
the scale of tens of meV. On experimental side, electron and hole
puddles were observed in scanning tunneling microscopy (STM) studies.
However, the interpretation of their origin is a matter of ongoing
discussion.\cite{desphande:prb2009,zhang:natphys2009}

While the qualitative estimate of scattering rates associated with
ripples were presented in Ref.~\onlinecite{katsnelson:pht08}, little
in a way of explicit calculations of transport properties is present
in the literature to date. Simple one-dimensional models were studied
in Refs.~\onlinecite{Isacsson:prb2008,zwierzycki:app2012}. More
sophisticated structural models were used in Refs.~\onlinecite{Klos:PRB2009}
and \onlinecite{Zhu:PhysLettA2013} but the effect of rippling was
found to be modest compared to scattering on charged impurities. In
both cases the modeling of the ripples was tailored to the specific
case of relatively smooth structures of Ref.~\onlinecite{fasolino:natm07},
characterized by $H=1$. The aim of the present contribution is to
study the transport properties for the structures in the whole range
of $H$ and $std(h)$ values. In the following paragraphs I will outline
the details of the method of calculations and then present the results
for both ballistic and diffusive regimes. Based on these I argue that
the scattering on ripples has measurable impact on transport properties
and leads in fact to resistance and mobility values compatible with
experimental data.

\begin{figure}
\includegraphics[width=1\columnwidth]{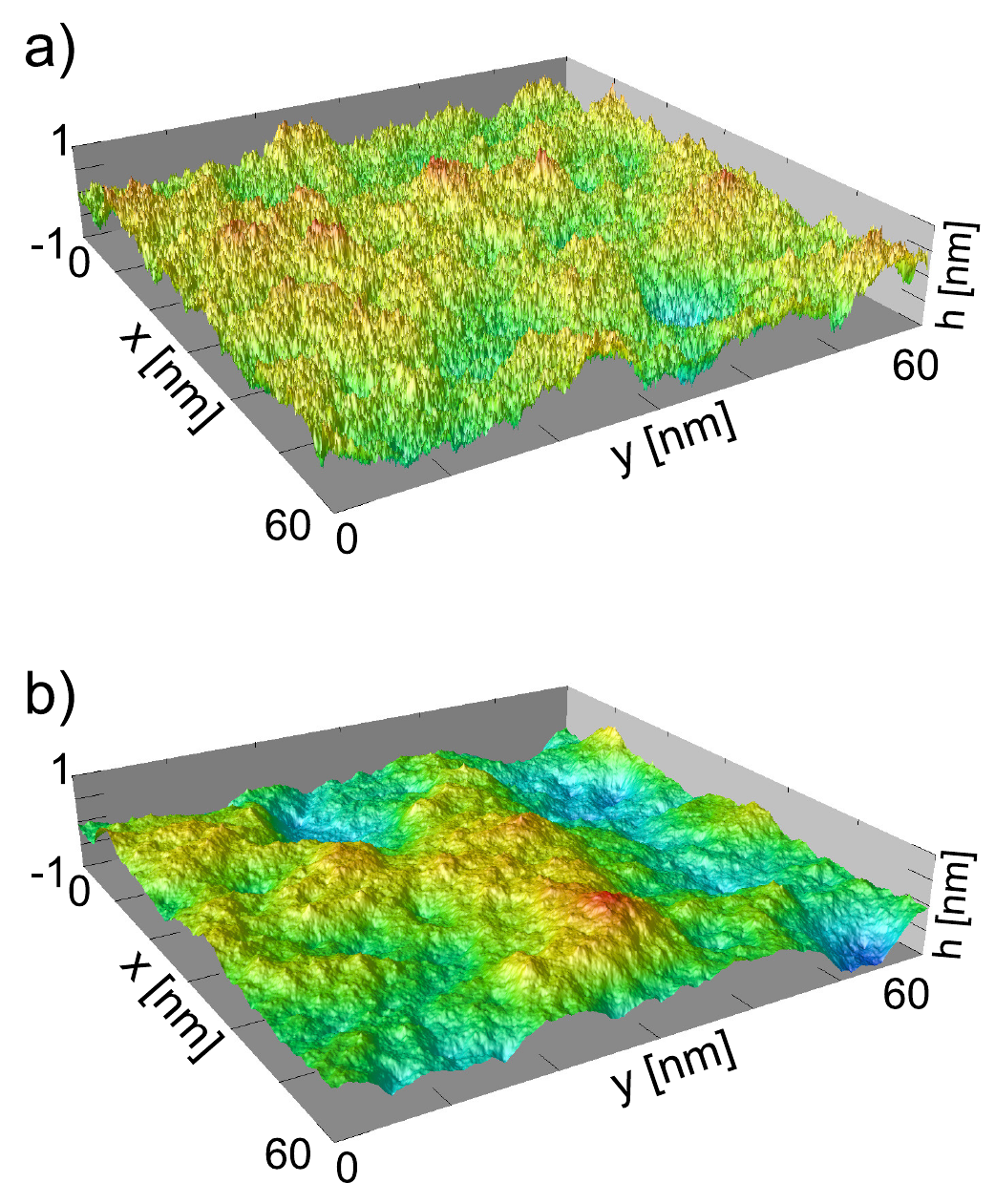}

\caption{Two examples of surfaces generated using diamond-square algorithm.
Scaling exponent $H$ was set to $0.5$ and $1.0$ in a) and b) panels
respectively. Standard deviation $std(h)=3$~\AA\ was used in both
cases.}
\label{fig:surf}
\end{figure}

\begin{figure}
\includegraphics[width=0.9\columnwidth]{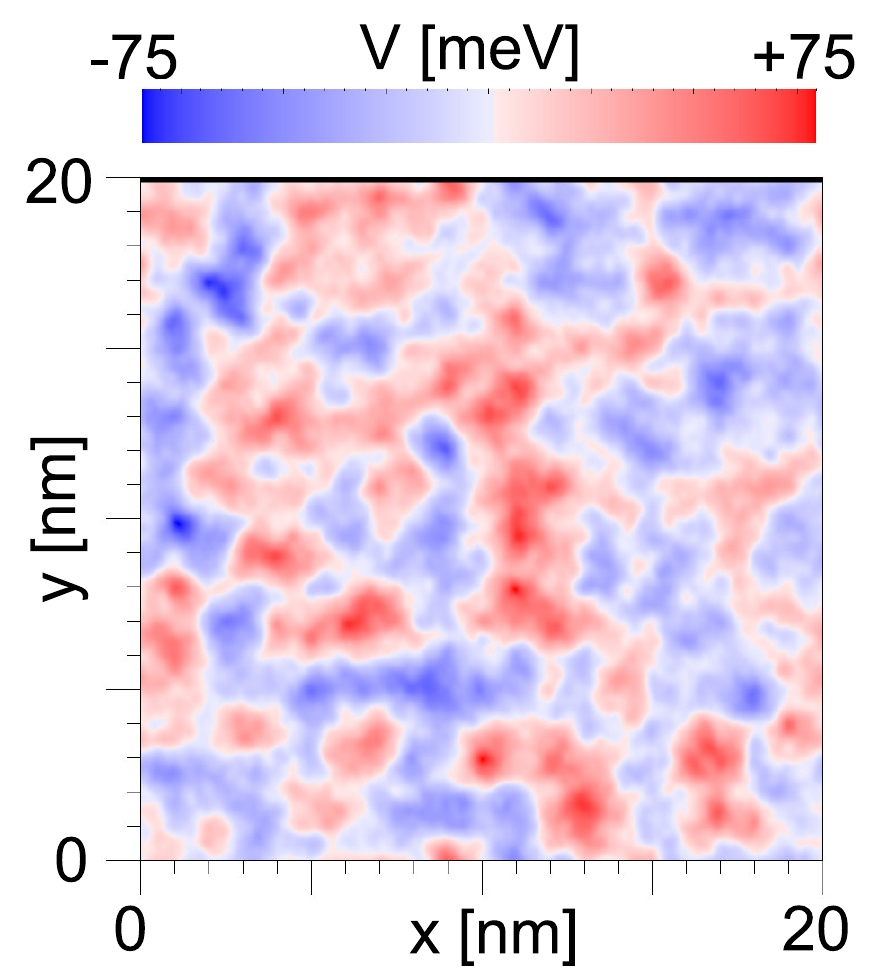}

\caption{Scalar potential generated using diamond-square algorithm with $H=1.0$
and $std(V_{s})=50\;\mathrm{meV}$ using tiles of $2\times2\;\mathrm{nm}$
(type A).}
\label{fig:scalar}
\end{figure}

\section{Method}

The calculations were performed using a single-band tight-binding
model with hopping restricted to nearest neighbors only. The effect
of rippling was included in the Hamiltonian by modification of the
hopping constants. A simple scaling formula was adopted to this end
\begin{equation}
t(r)=t_{0}\left(r/a_{0}\right)^{-3}\label{eq:cubic_scaling}
\end{equation}
where $r$ is the interatomic distance, $a_{0}=1.42$~\AA\ is equilibrium
lattice constant and $t_{0}=t(a_{0})$ equals $-2.8\,\mathrm{eV}$.
The choice of scaling formula is further discussed in Appendix~\ref{sec:The-Scaling-of}.
The scattering geometry was employed with system consisting of rippled
region sandwiched between flat-graphene leads. The leads were assumed
to be heavily doped with the Fermi energy located $1\,\mathrm{eV}$
over neutrality point. For such system transmission coefficients $t_{\mu\nu}$
between the incoming state $\nu$ in the left and outgoing state $\mu$
in the right electrode were calculated using wave function matching
method.\cite{Xia:prb06,Zwierzycki:pssb08} These were subsequently
used in the Landauer-B\"uttiker formula for conductance 
\begin{equation}
G_{LB}=\frac{e^{2}}{h}\sum_{\mu\nu}\left|t_{\mu\nu}\right|^{2}.\label{eq:landauer}
\end{equation}

The self-similar landscape of the scattering region was generated
using diamond-square algorithm,\cite{Fournier:CommACM1982} described
in the Appendix~\ref{sec:Diamond---Square}, as multiples of $20\times20\;\mathrm{nm}$
non-repeating tiles. This yields the correlation length of $\xi\approx10\;\mathrm{nm}$.
The exponent $H$ and $std(h)$ were adjustable parameters. Two examples
of generated surfaces are shown in Fig.~\ref{fig:surf}. The honeycomb
lattice of graphene was mapped onto the resulting surface with no
adjustments to the in-plane positions of carbon atoms. 

In some calculations a fluctuating scalar potential, included via
the on-site elements of the tight-binding Hamiltonian, was taken into
account. In contrast to Refs.~\onlinecite{gazit:prb(R)2009,gibertini:prb10,gibertini:prb(r)2012,Kim:EPL2008,partovi:prb2011}
where the calculated local potentials were explicitly dependent on
the geometry, a simplified approach was adopted here. The potentials
were generated independently, using the same algorithm as geometry,
with height replaced by local potential. Two different tile sizes
were employed, $2\times2$ and $20\times20\;\mathrm{nm}$, in order
to create the potentials varying over shorter (\emph{type A}) or longer
(\emph{type B}) length scales. The $H=1$ exponent and additional
averaging over neighbors was used so as to ensure the smoothness of
resulting potentials. An example is shown in Fig.~\ref{fig:scalar}. 

In all the calculations the transport was assumed to be along armchair
direction and periodic boundary conditions with period $W$ between
$40$ and $80\;\mathrm{nm}$ were used in the lateral (zigzag) direction.
All the results presented in the following section were averaged over
several tens realizations of the scattering region.

\begin{figure}
\noindent \includegraphics[width=0.88\columnwidth]{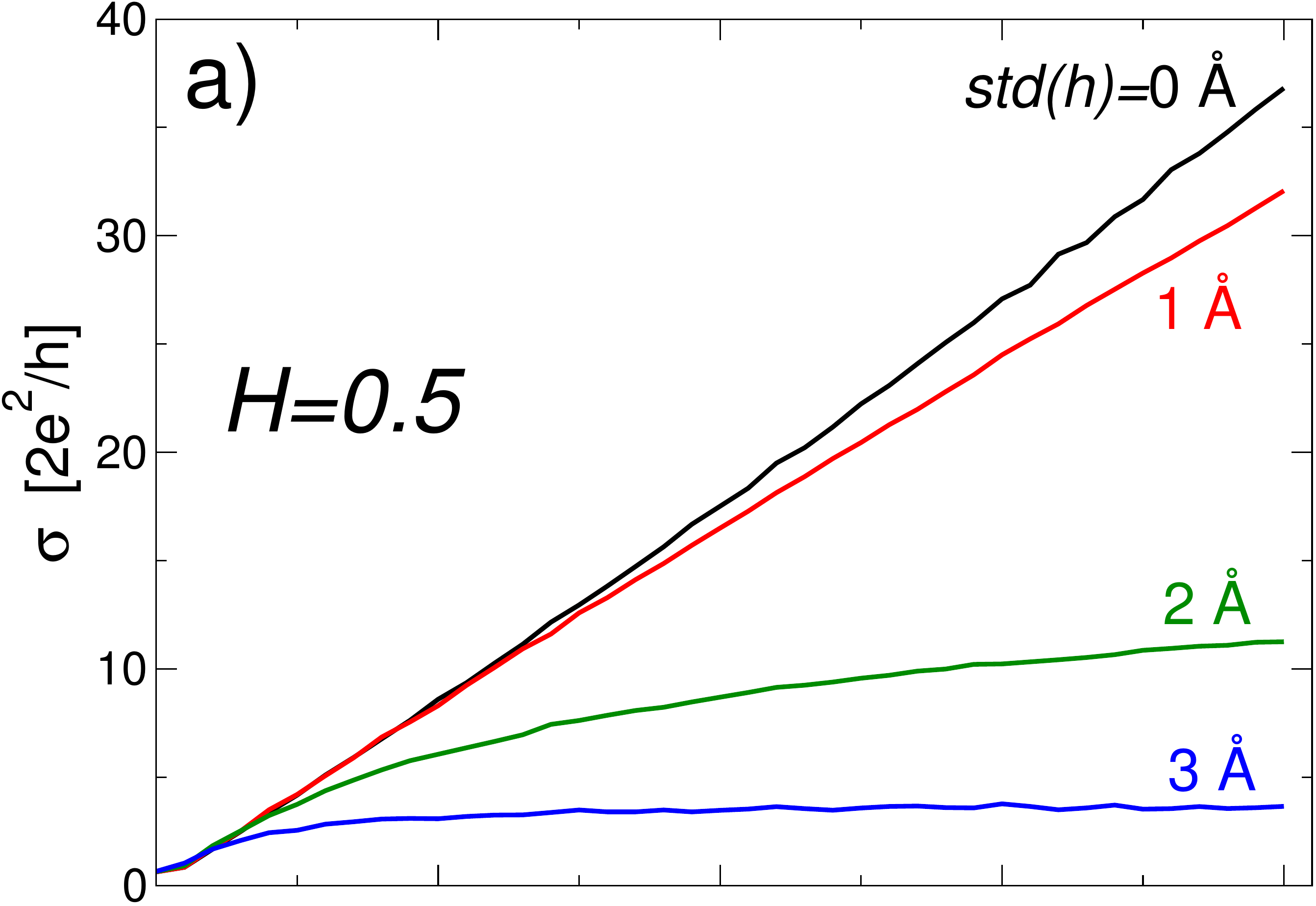}

\noindent \vskip0.1cm

\noindent \includegraphics[width=0.88\columnwidth]{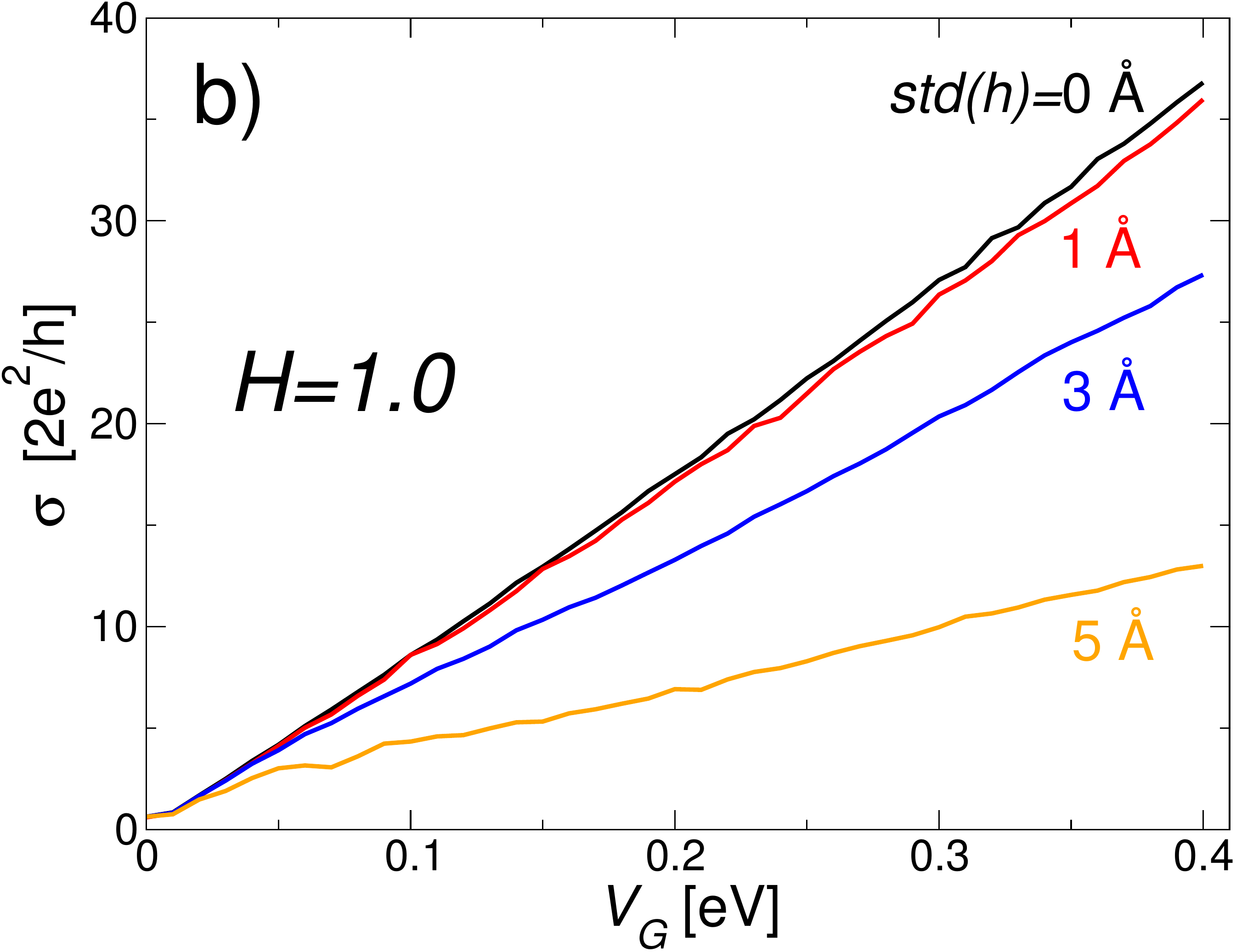}

\caption{Conductivity $\sigma=G_{LB}\frac{L}{W}$ of the scattering region
with $L=100\;\mathrm{nm}$ and $W=20\;\mathrm{nm}$ calculated as
a function of $V_{G}$, the distance between the Fermi level and the
neutrality point in the scattering region. Different curves correspond
to the different values of $std(h)$. }
\label{fig:cond_vg}
\end{figure}

\begin{figure}
\includegraphics[width=0.88\columnwidth]{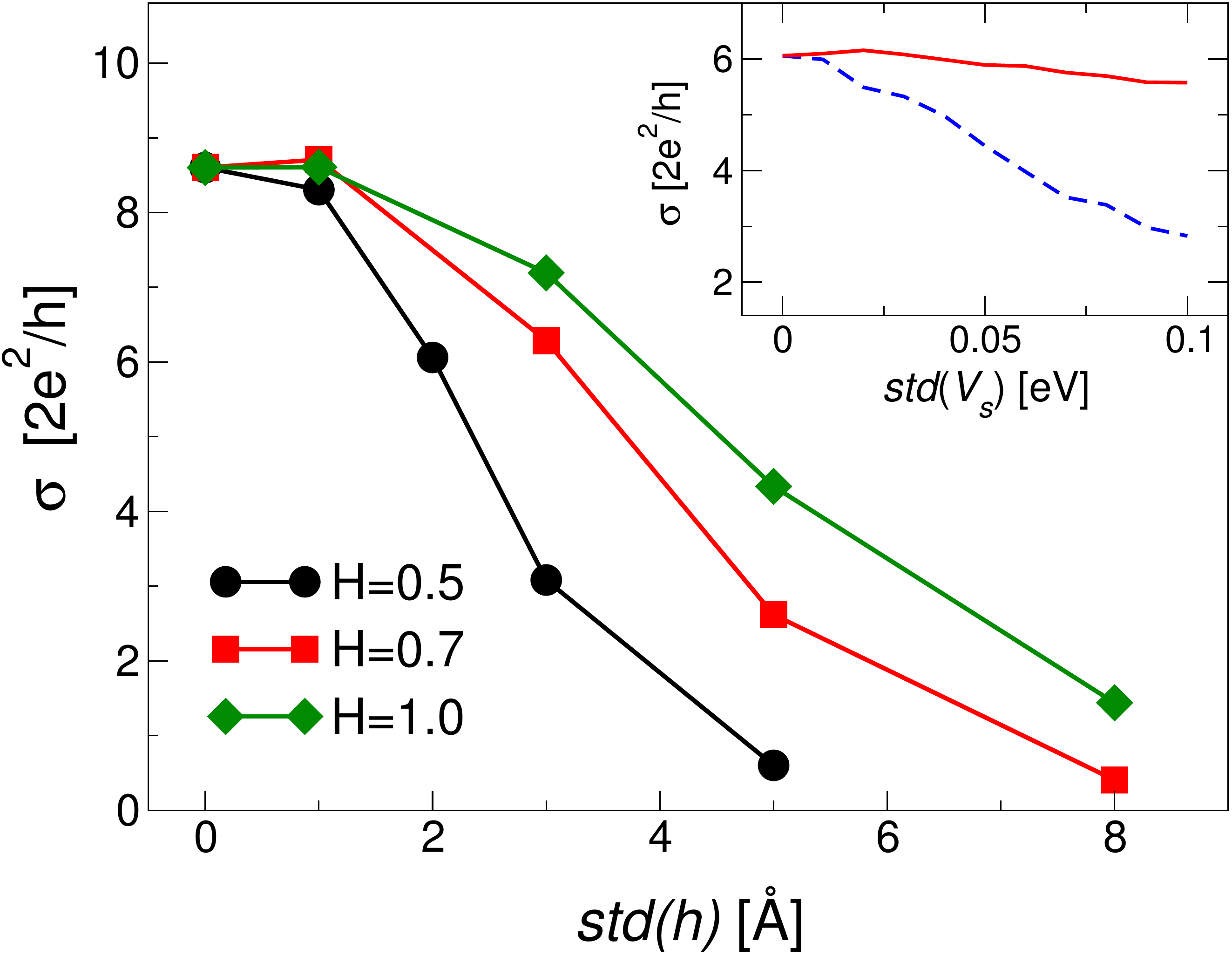}

\caption{Conductivity as function of $std(h)$ for fixed $V_{G}=0.1\,\mathrm{eV}$
and different values of scaling exponent $H$. \emph{The inset:} Conductivity
as function of the standard deviation of scalar potential $std(V_{s})$
for $H=0.5$, $std(h)=2$~\AA\ and $V_{G}=0.1\;\mathrm{eV}$. The
results for \textcolor{red}{\emph{type A}} (solid line) and \textcolor{blue}{\emph{type
B}} (dashed line) potentials are presented.}

\label{fig:sig_dep}
\end{figure}

\section{Results}

In Fig.~\ref{fig:cond_vg} the calculated conductivity $\sigma=G_{LB}\frac{L}{W}$
(for $L=100\;\mathrm{nm}$ and $W=20\;\mathrm{nm}$) is shown as a
function of a distance between the Fermi level and the neutrality
point%
\footnote{The value of $V_{G}$ is calculated assuming that rippling does not
change the position of neutrality point.%
} in the scattering region. This quantity, $V_{G}$, would be controlled
by gate voltage in experiment. The convention used here is that positive
values mean electron doping in the central part. It is immediately
apparent that the conductance can be significantly lowered by the
presence of the ripples. The effect increases for higher values of
$std(h)$. Additionally the functional dependence of $\sigma$ can
also change and deviate from $\sigma\sim V_{G}$, typical for ballistic
regime, as seen most clearly in Fig.~\ref{fig:cond_vg}a. Comparison
of results obtained for $H=0.5$ and $1.0$ reveals that the back-scattering
is greatly reduced in the latter case. This correlates with the structures
characterized by higher $H$ values being locally smoother (see Fig.~\ref{fig:surf}).
This point is further illustrated in Fig.~\ref{fig:sig_dep} were
the conductivities for three values of $H$ are shown, for a fixed
value of $V_{G}=0.1\,\mathrm{eV}$ (corresponding to $n=10^{12}\,\mathrm{cm^{-2}}$
in flat graphene), as function of $std(h)$. The descent of the curves
gets slower with increasing $H$, however even for $H=1$, the effect
is non-negligible. The dependence of conductivity on the standard
deviation of the scalar potential ,$std(V_{s})$, is presented in
the inset of Fig.~\ref{fig:sig_dep} for $H=0.5$ and $std(h)=2$\AA.
One sees that of the two model potentials, the ``long-wavelength''
type B has significantly stronger effect.

\begin{figure}[t]
\includegraphics[width=1\columnwidth]{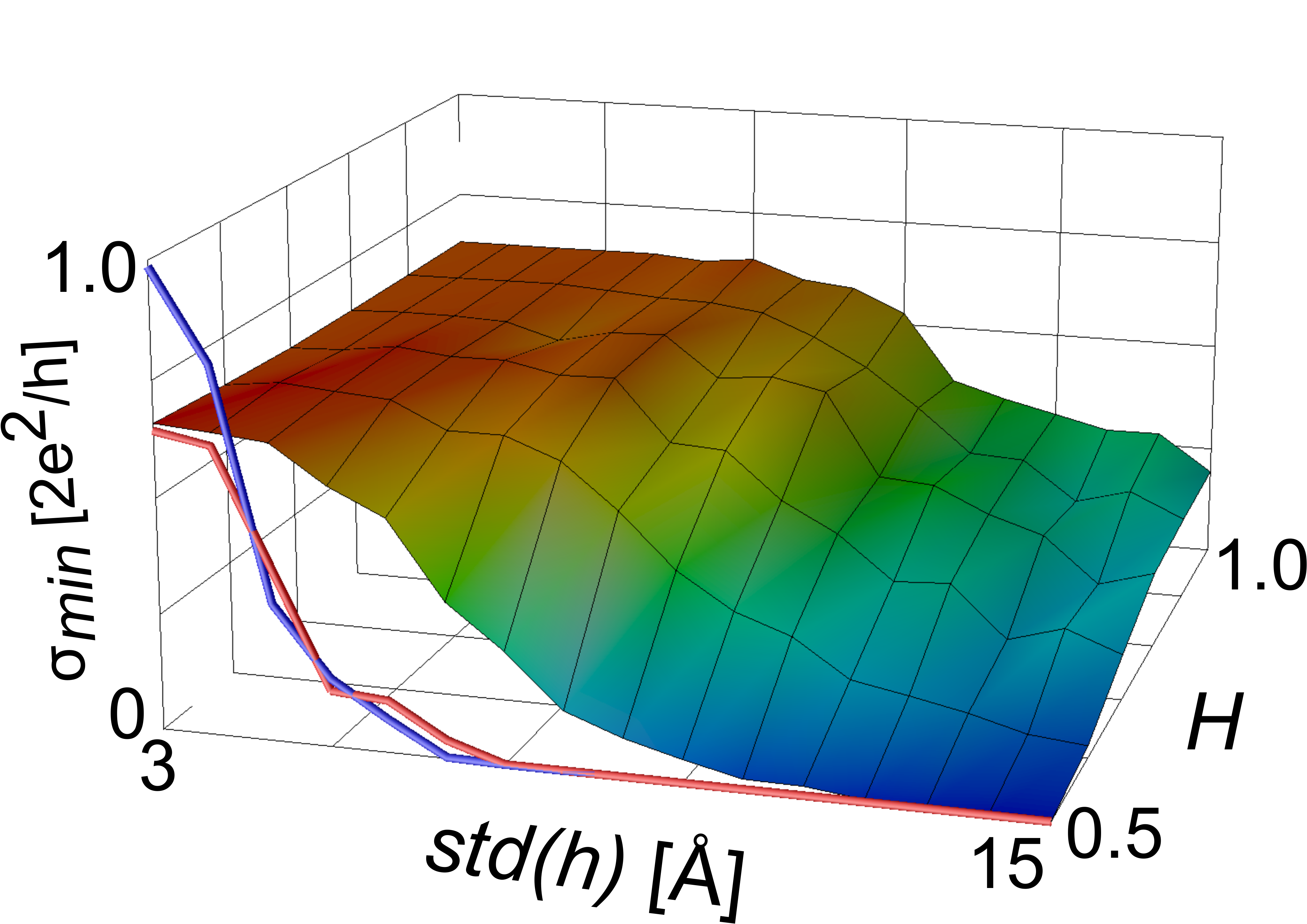}

\caption{Minimal conductivity $\sigma_{min}=\sigma(V_{G}=0)$ plotted as function
of $H$ and $std(h)$. The plateau values correspond to the universal
value of $\sigma_{min}\approx\frac{4e^{2}}{h\pi}$. No scalar potential
was present for the main surface plot. The results including fluctuating
scalar potential were shown as red and blue lines for \textcolor{red}{\emph{type
A}} and \textcolor{blue}{\emph{type B}} potentials (with $std(V_{s})=0.05\;\mathrm{eV}$
and $H=0.5$) respectively.}
\label{fig:min_cond}
\end{figure}

Interestingly, the region around neutrality point ($V_{G}\approx0$)
in Fig.~\ref{fig:cond_vg} seems not to be affected by rippling.
This is confirmed in Fig.~\ref{fig:min_cond} where the minimal conductance,
$\sigma_{min}$=$\sigma(V_{G}=0)$ is shown as a function of $H$
and $std(h)$. The main surface plot corresponds to the case with
no additional scalar potential. We observe that $\sigma_{min}$ assumes
the values close to the \emph{universal} value of $\frac{4e^{2}}{h\pi}$,
predicted for flat graphene,\cite{Tworzydlo:prl2006} over large plateau
and decreases only for $std(h)$ exceeding $H$-dependent threshold
value. Increasing the length of the scattering region ($L$) yields
essentially same plateau region (results not shown) with a steeper
descent outside. Adding the \emph{type A} scalar potential does not,
for $std(V_{s})\lesssim0.1\,\mathrm{eV}$, change the value of minimal
conductance but reduces the size of the plateau. Exemplary results
are shown in Fig.~\ref{fig:min_cond} using the red line for $H=0.5$
and $std(V_{s})=0.05\,\mathrm{eV}$. The \emph{type B} potential,
which varies laterally on the same length scale as structural corrugations,
is capable of changing not only stability but also the value of minimal
conductance as demonstrated by the blue curve in Fig.~\ref{fig:min_cond}
(same parameters apply). The observed increase of $\sigma_{min}$
in the presence of the slow varying scalar potential can be qualitatively
understood by noticing that in this case a carrier travels through
a series of locally conductive regions (\emph{i.e.} electron and holes
puddles). For specific combinations of $H$, $std(h)$ and $std(V_{s}$),
$\sigma_{min}$ can achieve a value of $\frac{4e^{2}}{h}$ often seen
in exfoliated samples.\cite{Novoselov:nat05} It should be noted however,
that such values are no longer universal in my calculations, \emph{i.e.}
in addition to above mentioned variables they also depend on sample
dimensions.

\begin{figure}[t]
\includegraphics[width=0.9\columnwidth]{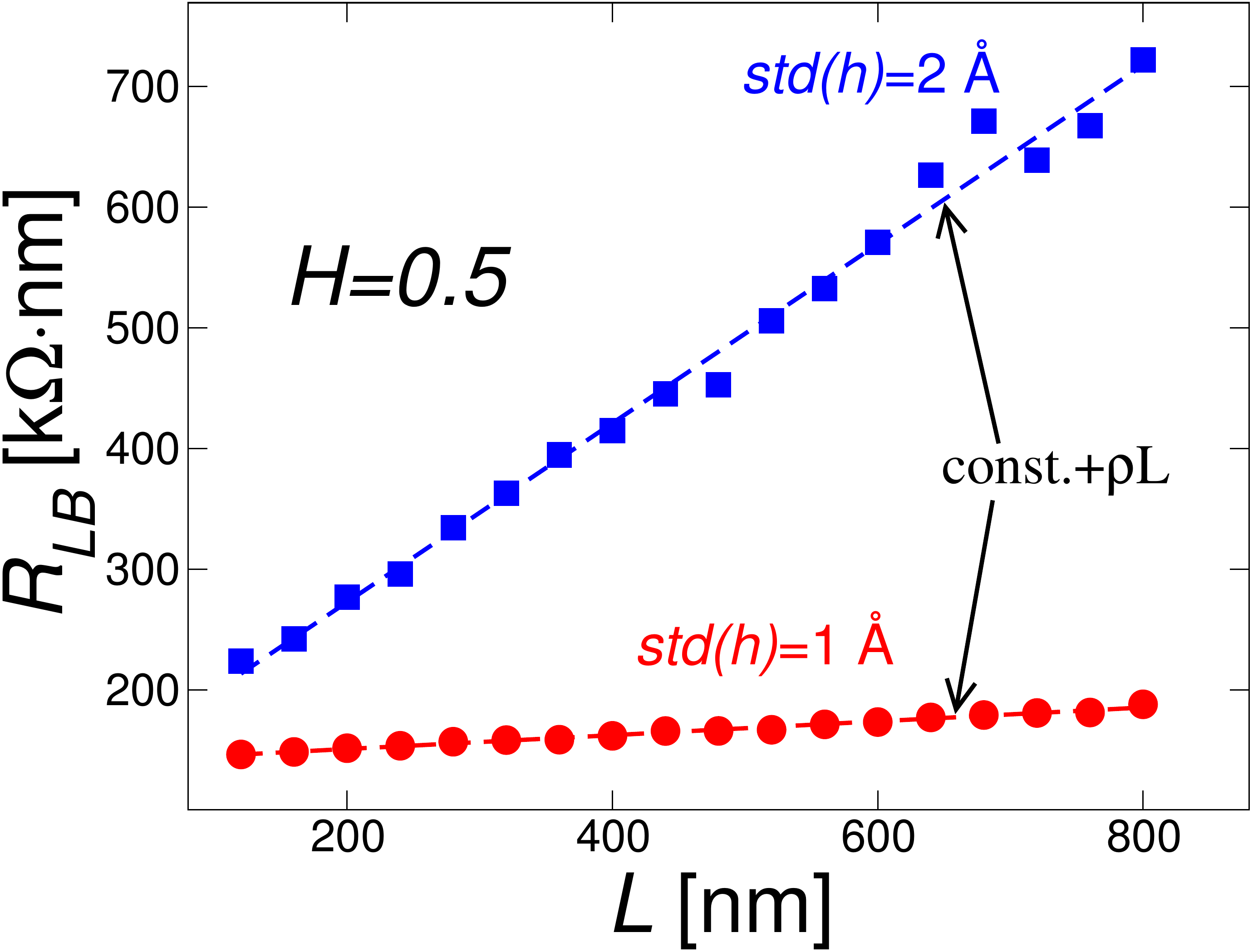}

\caption{The illustration of the fitting procedure described in the text. The
points on the plot are calculated values of $R_{LB}=G_{LB}^{-1}$
for $H=0.5$ and $std(h)$ equal to either $1$~\AA\ (\textcolor{red}{red
circles}) or $2$~\AA\ (\textcolor{blue}{blue squares}) respectively.
The dashed lines are the results of the least square fit, as described
by Eq.~(\ref{eq:fit}).}
\label{fig:fit}
\end{figure}

\begin{figure}
\includegraphics[width=1\columnwidth]{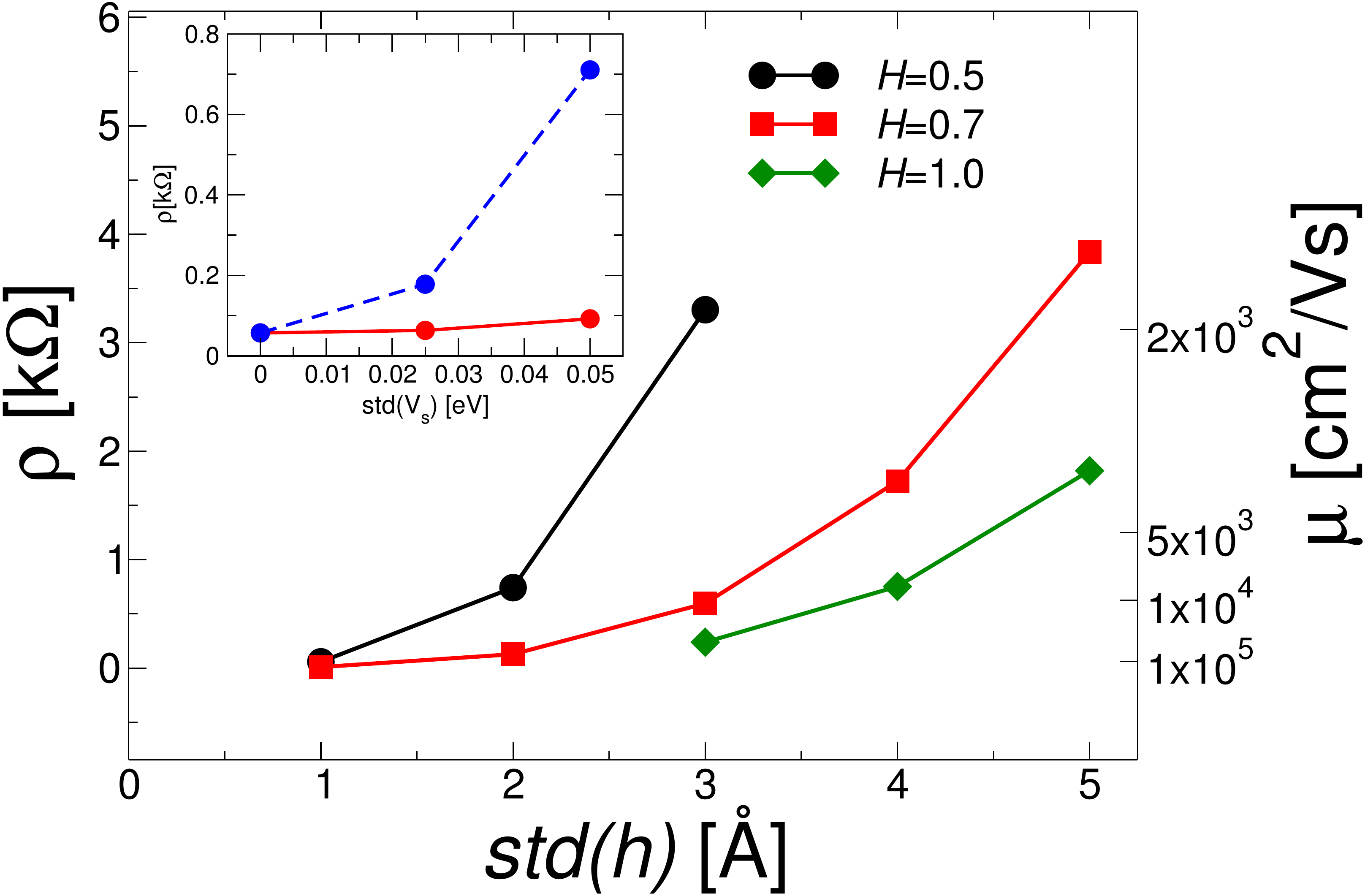}

\caption{Resistivity calculated using Eq.~(\ref{eq:fit}) for different values
of roughness exponent $H$, as function of $std(h)$. $V_{G}=0.1$~\AA\
was used in all cases. The lines are guides to the eye only. The values
of the corresponding mobilities can be read using the scale on the
right vertical axis. \emph{The inset:} Resistivity as function of
$std(V_{s})$ calculated with the inclusion of \textcolor{red}{\emph{type~A}}
(solid line) and \textcolor{blue}{\emph{type~B}} (dashed line) scalar
potentials potentials. $H=0.5$ and $std(h)=1$~\AA\ was used for
both cases.}
\label{fig:rho}
\end{figure}

Following the usual convention the results of Figs.~\ref{fig:cond_vg}
and \ref{fig:min_cond} were presented as \emph{conductivit}ies, $\sigma$.
It should be understood, however, that these results characterize
a specific sample of given dimensions rather than bulk of a material.
In fact, a naive attempt to extract conductivity or resistivity, as
featured in Ohms law, from Eq.~(\ref{eq:landauer}) is destined to
fail as Landauer-B\"uttiker formula yields a finite resistance of
a scattering region even in $L=0$ limit.\cite{Sharvin:zetf65,Datta:95}
On the other hand one expects that if the scattering region is sufficiently
disordered, the overall resistance calculated as inverse of Eq.~(\ref{eq:landauer})
should contain a contribution which scales linearly with $L$.\cite{xu:prb2006}
Therefore by calculating $G_{LB}^{-1}(L)$ and performing a least
square fit to
\begin{equation}
R_{LB}=G_{LB}^{-1}(L)=const.+\rho\cdot L\label{eq:fit}
\end{equation}
it is possible to estimate resistance appropriate for ohmic regime.
A constant term is, in present case, expected to be close to the inverse
of Sharvin conductance,\cite{Sharvin:zetf65,Datta:95} $G_{Sh}=\frac{e^{2}}{h}N$,
proportional to the number of current currying modes in the scattering
region. The procedure is illustrated in Fig.~\ref{fig:fit} for $H=0.5$
and $std(h)=1$ and $2$~\AA. The resistivities obtained using Eq.~(\ref{eq:fit})
and $R_{LB}(L)$ calculated for lengths of up to $L=800\;\mathrm{nm}$,
are shown in Fig.~\ref{fig:rho}. No calculations were performed
for $H=1.0$ and $sdt(h)<3$\AA\ because the weak scattering in this
case makes it difficult to reach the linear scaling regime. Once again
the potential shift of the scattering region was set to $V_{G}=0.1\,\mathrm{eV}$.
As expected we observe that resistivity increases with $std(h)$ but
decreases with growing $H$ in qualitative agreement with Fig.~\ref{fig:cond_vg}.
The inset of Fig.~\ref{fig:rho} shows the dependence of resistivity
(for $std(h)=1$~\AA\ and $H=0.5$) on the magnitude {[}$std(V_{s})${]}
of the scalar potential. We observe that additional scattering on
fluctuating scalar potential can substantially increase resistivity,
especially in the case of slow varying \emph{type B} potential. Only
very modest increased is seen for short-wavelength \emph{type A} potential.
This situation is analogous to the results for ballistic regime shown
in the inset of Fig.~\ref{fig:sig_dep}. The range of calculated
values overlaps with the experimentally observed range.\cite{Novoselov:sc04,Bolotin:prl2008,Du:NNano2008,ponomarenko:prl2009} 

Having calculated resistivity we can now estimate the mobilities,
$\mu$, corresponding to Fig.~\ref{fig:rho}. Using $\rho^{-1}=en\mu$
and assuming $n=10^{12}\,\mathrm{cm^{-2}}$, appropriate for flat
graphene with a chosen $V_{G}$, we obtain the values of $\mu$ which
can be read from the scale on the right vertical axis in Fig.~\ref{fig:rho}.
The STM studies reveal that exfoliated graphene on SiO$_{2}$ substrate
is characterized by $H\approx0.5$ and $std(h)\approx2-3$~\AA. This
range of parameters corresponds in my calculations to the mobility
in $\mu=2.000-8.000\;\mathrm{\frac{cm^{2}}{V\cdot s}}$ range which
is somewhat less than experimental values, typically equal to $10.000-20.000\;\mathrm{\frac{cm^{2}}{V\cdot s}}$.\cite{Novoselov:nat05}
The values of mobility are even more underestimated for the range
of parameters corresponding to suspended graphene flakes, where simulations\cite{fasolino:natm07,Los:prb2009,Abdepour:prb2007}
predict $H\approx0.7-1.0$ and $std(h)$ of several Angstroms. This
puts the calculated mobilities (see corresponding curves in Fig.~\ref{fig:rho})
in the $10^{3}\;\mathrm{\frac{cm^{2}}{V\cdot s}}$ range, whereas
much higher values of $100.000\mathrm{\;\frac{cm^{2}}{V\cdot s}}$
and more were reported experimentally.\cite{Bolotin:prl2008,Du:NNano2008}.
One possible reason for overestimation of scattering strength in my
calculations may be the lack of structural relaxation in the model.
It has been in fact demonstrated in the literature that relaxation
of internal strain leads to the substantial reduction of the effect
of rippling on electronic structure of the sample.\cite{wehling:epl08} 

\section{Conclusions}

In this paper I have studied the transport properties of rippled graphene
using structural model parameterised by the scaling exponent $H$
and the standard deviation $std(h)$ of the out-of plane positions
of carbon atoms. Both in ballistic and ohmic regime a substantial
reduction of conductivities was found already for relatively modest
amplitudes of corrugation. The range of resistivities and mobilities
calculated in ohmic regime overlaps with experimental values. The
mobility is however consistently underestimated when direct comparison
is being made, possibly because of the limitations of the structural
model. The value of minimal conductivity was found to be largely unchanged
and equal to universal value $\sigma_{min}=\frac{4e^{2}}{h\pi}$ when
only the effect of changing bond lengths was considered. The fluctuating
scalar potential was found to decrease conductance for doped samples
and increase the minimal conductance for globally neutral samples.
In conclusion, the results presented in this paper support the notion
that the rippled induced scattering is one of the important factors
limiting the mobility of carriers in graphene.
\begin{acknowledgments}
This work was supported by the Polish Ministry of Science and Higher
Education as a research project No. N~N202~199239. 
\end{acknowledgments}
\appendix

\section{The Scaling of The Hopping Constant\label{sec:The-Scaling-of}}

A simple formula for the scaling of the hopping constant used throughout
the paper
\begin{equation}
t(r)=t_{0}\left(\frac{r}{a_{0}}\right)^{-3}\label{eq:lmto_scaling}
\end{equation}
is a special case of a general $t_{ll'}\sim r^{-l-l'-1}$ scaling
(with $l=l'=1$ for graphene $\pi$-band) motivated in turn by the
scaling properties of the Hamiltonian of Linear Muffin Tin Orbital
(LMTO)\cite{Andersen:85} method. Other choices, frequently utilized
in the literature, include a quadratic scaling suggested by Harrison
\cite{harrison:elementary2004,*[{A generalized version of Harrison's scaling suggested in }][{ does not bring consistent improvement in the present case.}]xu:jphys1992}
\begin{equation}
t(r)=t_{0}\left(\frac{r}{a_{0}}\right)^{-2}\label{eq:harrison}
\end{equation}
or a more involved formula of Pereira \emph{et al.\cite{pereira:prb2009}
\begin{equation}
t(r)=t_{0}e^{-3.37\left(\frac{r}{a_{0}}-1\right)}\label{eq:pereira}
\end{equation}
}

\begin{figure}[t]
\includegraphics[width=0.8\columnwidth]{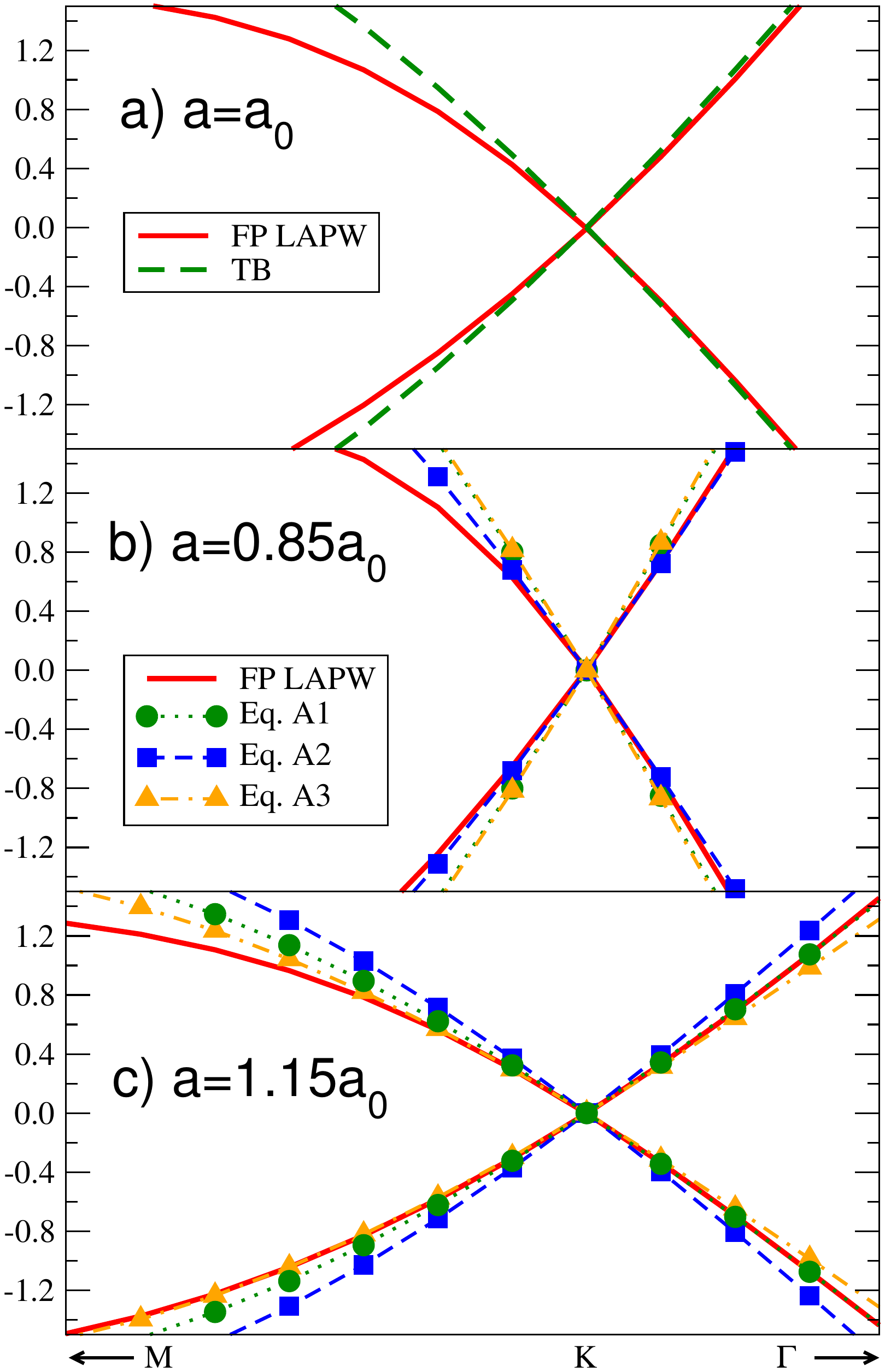}

\caption{Graphene band structure around K-point calculated using tight-binding
(TB) model and \emph{ab initio} FLAPW method with lattice constant
being a) at equilibrium value; b) contracted and c) expanded. The
results of the three scaling formulas discussed in the Appendix~\ref{sec:The-Scaling-of}
concide in panel a) and are marked using circles (\CIRCLE{}), squares
($\blacksquare$) and triangles ($\blacktriangle$) for Eq.~(\ref{eq:lmto_scaling}),
(\ref{eq:harrison}) and (\ref{eq:pereira}) respectively.}
\label{fig:scaling}
\end{figure}

The performance of all three scaling formulas is compared in Fig.~\ref{fig:scaling}
for both contracted (middle panel) and expanded (bottom panel) lattice
constant. Additionally the results of tight-binding (TB) calculations
were benchmarked against \emph{ab-initio} Full Potential Linearized
Augmented Plane Wave Method (FP~LAPW).\cite{[{As inplemented in WIEN2k program: }]Schwarz:ComPhysComm2002}
From the results shown in Fig.~\ref{fig:scaling} its is clear that
none of the formulas is consistently better than the other two. In
the case of contracted lattice constant the best agreement between
\emph{ab intio} and TB results is given by Eq.~(\ref{eq:harrison})
with the other two being indistinguishable. However, the situation
is reversed for expanded lattice constant when the \emph{ab initio}
bands are best reproduced by either Eq.~(\ref{eq:pereira}) (along
$M-K$ direction) or Eq.~(\ref{eq:lmto_scaling}) (along $K-\Gamma$)
with Eq.~(\ref{eq:harrison}) clearly the worst of the three.

In the effective mass approximation, as used \emph{e.g.} in Refs.
\onlinecite{guinea:prb08a,guinea:prb08b,vozmediano:physrep2010},
a linear scaling of the hopping constant is assumed and only the slope,
usually parameterized as logarithmic derivative
\[
\beta=-\left.\frac{d\ln t}{d\ln r}\right|_{r=a_{0}},
\]
enters the theory. In this picture the Eqs.~(\ref{eq:lmto_scaling}),
(\ref{eq:harrison}), (\ref{eq:pereira}) correspond to $\beta$ equal
to 3, 2 and 3.37 respectively. The theoretical and experimental estimates
of $\beta$ values found in the literature range between 1.1 and 3.6.\cite{suzuura:prb2002,*castro-neto:prb2007,*[{}][{ and references therein.}]cappelluti:prb2012}

\begin{figure}[t]
\includegraphics[width=0.8\columnwidth]{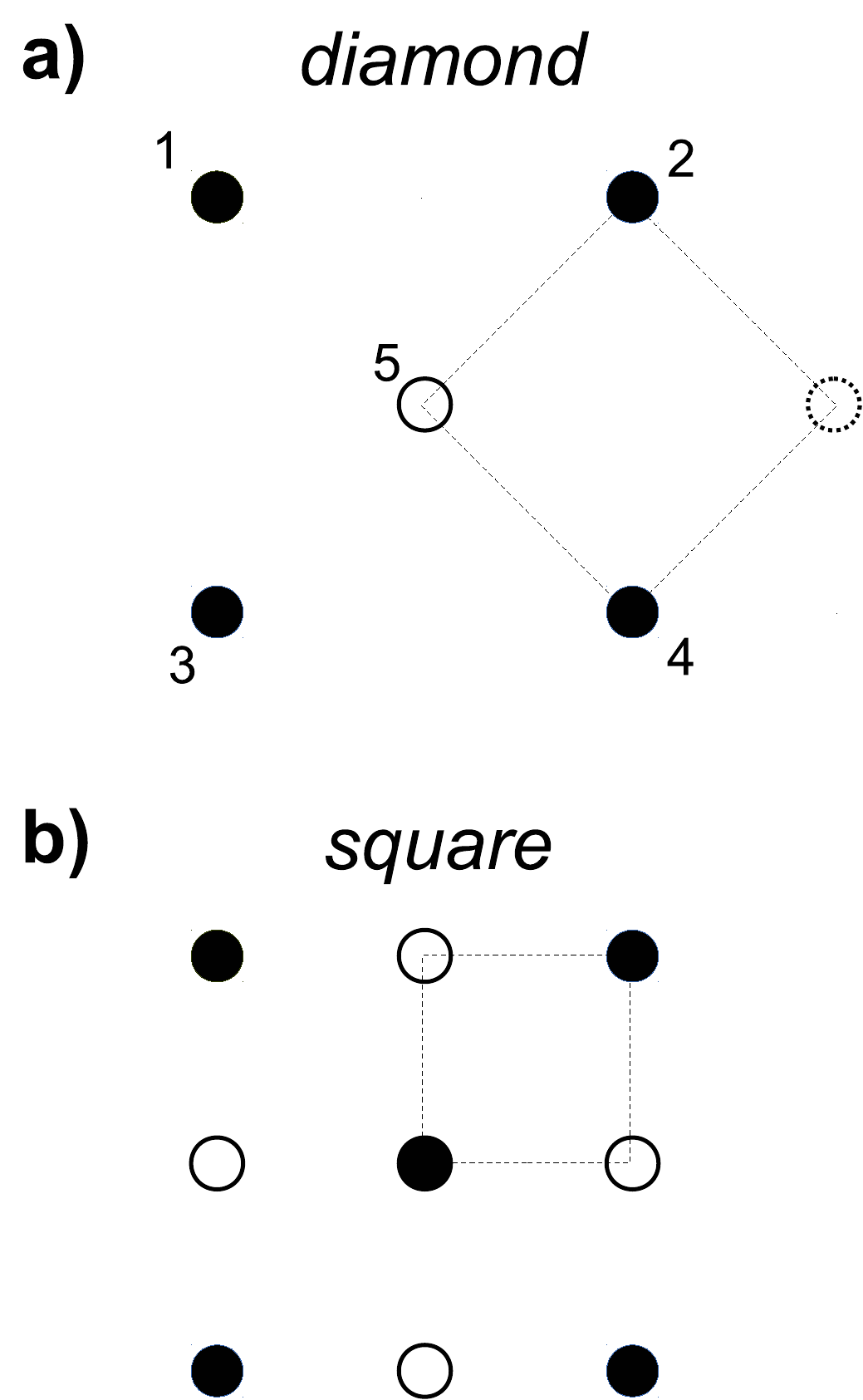}

\caption{The two steps of diamond-square algorithm. The full circles ($\bullet$)
indicate points which values are known at the beginning of a step.
The values of the points marked with empty circles ($\circ$) are
calculated during the step. The dashed figures explain the name of
each step.}
\label{fig:diamond}
\end{figure}

\section{Diamond -- Square Algorithm}

\label{sec:Diamond---Square}The diamond-square algorithm, first proposed
in Ref.~\onlinecite{Fournier:CommACM1982}, is an iterative procedure
for generating fractal landscapes with chosen value of scaling exponent
$H$. Consider a square with some predefined (possibly zero) corner
values. The first iteration, shown in Fig.~\ref{fig:diamond}, consists
of two steps:
\begin{itemize}
\item \emph{diamond:} the central value of the square is calculated as
\[
h_{5}=\sum_{i=1\ldots4}h_{i}+rand(d)
\]
where $rand(d)$ is a random variable with normal distribution%
\footnote{Alternatively a uniform distribution over \unexpanded{$[-d,d]$} can
be used.%
} centered at zero and standard deviation equal to $d$,
\item \emph{square:} here one calculates the central values of the diamonds
in analogous fashion.
\end{itemize}
Cyclic boundary conditions can be used in the first iteration or alternatively
one can skip the points outside of the original rectangle. Both steps
are repeated required number of times for squares and diamonds with
dimensions halved in each iteration. Importantly, after each iteration
the standard deviation of the random variable is reduced according
to$ $
\[
d=\frac{d}{2^{H}}
\]
thus enforcing required scaling properties. The scaling exponent $H$
is related to the fractal dimension $D$ of the structure via $D=3-H$.

It is easy to generalize the algorithm so that it works with rectangles
consisting of multiple square tiles. This is the approach adopted
in the paper, with the size of a basic tile equal to $20\times20\;\mathrm{nm}$.
Thus obtained set of $h$ was renormalized so as to ensure $\bar{h}=0$
and desired value of the standard deviation $std(h)$. The number
of iterations were set to 8 resulting in a square grid with neighboring
points being about $0.7$~\AA\ apart, that is approximately half
of interatomic distance in graphene. The hexagonal lattice of graphene
was then overlaid on the square-grid surface by setting appropriate
heights. The in-plane coordinates were not modified.


\begin{thebibliography}{56}%
\makeatletter
\providecommand \@ifxundefined [1]{%
 \@ifx{#1\undefined}
}%
\providecommand \@ifnum [1]{%
 \ifnum #1\expandafter \@firstoftwo
 \else \expandafter \@secondoftwo
 \fi
}%
\providecommand \@ifx [1]{%
 \ifx #1\expandafter \@firstoftwo
 \else \expandafter \@secondoftwo
 \fi
}%
\providecommand \natexlab [1]{#1}%
\providecommand \enquote  [1]{``#1''}%
\providecommand \bibnamefont  [1]{#1}%
\providecommand \bibfnamefont [1]{#1}%
\providecommand \citenamefont [1]{#1}%
\providecommand \href@noop [0]{\@secondoftwo}%
\providecommand \href [0]{\begingroup \@sanitize@url \@href}%
\providecommand \@href[1]{\@@startlink{#1}\@@href}%
\providecommand \@@href[1]{\endgroup#1\@@endlink}%
\providecommand \@sanitize@url [0]{\catcode `\\12\catcode `\$12\catcode
  `\&12\catcode `\#12\catcode `\^12\catcode `\_12\catcode `\%12\relax}%
\providecommand \@@startlink[1]{}%
\providecommand \@@endlink[0]{}%
\providecommand \url  [0]{\begingroup\@sanitize@url \@url }%
\providecommand \@url [1]{\endgroup\@href {#1}{\urlprefix }}%
\providecommand \urlprefix  [0]{URL }%
\providecommand \Eprint [0]{\href }%
\providecommand \doibase [0]{http://dx.doi.org/}%
\providecommand \selectlanguage [0]{\@gobble}%
\providecommand \bibinfo  [0]{\@secondoftwo}%
\providecommand \bibfield  [0]{\@secondoftwo}%
\providecommand \translation [1]{[#1]}%
\providecommand \BibitemOpen [0]{}%
\providecommand \bibitemStop [0]{}%
\providecommand \bibitemNoStop [0]{.\EOS\space}%
\providecommand \EOS [0]{\spacefactor3000\relax}%
\providecommand \BibitemShut  [1]{\csname bibitem#1\endcsname}%
\let\auto@bib@innerbib\@empty
\bibitem [{\citenamefont {Novoselov}\ \emph {et~al.}(2004)\citenamefont
  {Novoselov}, \citenamefont {Geim}, \citenamefont {Morozov}, \citenamefont
  {Jiang}, \citenamefont {Zhang}, \citenamefont {Dubonos}, \citenamefont
  {Grigorieva},\ and\ \citenamefont {Firsov}}]{Novoselov:sc04}%
  \BibitemOpen
  \bibfield  {author} {\bibinfo {author} {\bibfnamefont {K.~S.}\ \bibnamefont
  {Novoselov}}, \bibinfo {author} {\bibfnamefont {A.~K.}\ \bibnamefont {Geim}},
  \bibinfo {author} {\bibfnamefont {S.~V.}\ \bibnamefont {Morozov}}, \bibinfo
  {author} {\bibfnamefont {D.}~\bibnamefont {Jiang}}, \bibinfo {author}
  {\bibfnamefont {Y.}~\bibnamefont {Zhang}}, \bibinfo {author} {\bibfnamefont
  {S.~V.}\ \bibnamefont {Dubonos}}, \bibinfo {author} {\bibfnamefont {I.~V.}\
  \bibnamefont {Grigorieva}}, \ and\ \bibinfo {author} {\bibfnamefont {A.~A.}\
  \bibnamefont {Firsov}},\ }\href {\doibase 10.1126/science.1102896} {\bibfield
   {journal} {\bibinfo  {journal} {Science}\ }\textbf {\bibinfo {volume}
  {306}},\ \bibinfo {pages} {666} (\bibinfo {year} {2004})}\BibitemShut
  {NoStop}%
\bibitem [{\citenamefont {Novoselov}\ \emph {et~al.}(2005)\citenamefont
  {Novoselov}, \citenamefont {Geim}, \citenamefont {Morozov}, \citenamefont
  {Jiang}, \citenamefont {Katsnelson}, \citenamefont {Grigorieva},
  \citenamefont {Dubonos},\ and\ \citenamefont {Firsov}}]{Novoselov:nat05}%
  \BibitemOpen
  \bibfield  {author} {\bibinfo {author} {\bibfnamefont {K.~S.}\ \bibnamefont
  {Novoselov}}, \bibinfo {author} {\bibfnamefont {A.~K.}\ \bibnamefont {Geim}},
  \bibinfo {author} {\bibfnamefont {S.~V.}\ \bibnamefont {Morozov}}, \bibinfo
  {author} {\bibfnamefont {D.}~\bibnamefont {Jiang}}, \bibinfo {author}
  {\bibfnamefont {M.~I.}\ \bibnamefont {Katsnelson}}, \bibinfo {author}
  {\bibfnamefont {I.~V.}\ \bibnamefont {Grigorieva}}, \bibinfo {author}
  {\bibfnamefont {S.~V.}\ \bibnamefont {Dubonos}}, \ and\ \bibinfo {author}
  {\bibfnamefont {A.~A.}\ \bibnamefont {Firsov}},\ }\href {\doibase
  10.1038/nature04233} {\bibfield  {journal} {\bibinfo  {journal} {Nature}\
  }\textbf {\bibinfo {volume} {438}},\ \bibinfo {pages} {197} (\bibinfo {year}
  {2005})}\BibitemShut {NoStop}%
\bibitem [{\citenamefont {Castro~Neto}\ \emph {et~al.}(2009)\citenamefont
  {Castro~Neto}, \citenamefont {Guinea}, \citenamefont {Peres}, \citenamefont
  {Novoselov},\ and\ \citenamefont {Geim}}]{Neto:rmp08}%
  \BibitemOpen
  \bibfield  {author} {\bibinfo {author} {\bibfnamefont {A.~H.}\ \bibnamefont
  {Castro~Neto}}, \bibinfo {author} {\bibfnamefont {F.}~\bibnamefont {Guinea}},
  \bibinfo {author} {\bibfnamefont {N.~M.~R.}\ \bibnamefont {Peres}}, \bibinfo
  {author} {\bibfnamefont {K.~S.}\ \bibnamefont {Novoselov}}, \ and\ \bibinfo
  {author} {\bibfnamefont {A.~K.}\ \bibnamefont {Geim}},\ }\href {\doibase
  10.1103/RevModPhys.81.109} {\bibfield  {journal} {\bibinfo  {journal} {Rev.
  Mod. Phys.}\ }\textbf {\bibinfo {volume} {81}},\ \bibinfo {pages} {109}
  (\bibinfo {year} {2009})}\BibitemShut {NoStop}%
\bibitem [{\citenamefont {Bolotin}\ \emph {et~al.}(2008)\citenamefont
  {Bolotin}, \citenamefont {Sikes}, \citenamefont {Hone}, \citenamefont
  {Stormer},\ and\ \citenamefont {Kim}}]{Bolotin:prl2008}%
  \BibitemOpen
  \bibfield  {author} {\bibinfo {author} {\bibfnamefont {K.~I.}\ \bibnamefont
  {Bolotin}}, \bibinfo {author} {\bibfnamefont {K.~J.}\ \bibnamefont {Sikes}},
  \bibinfo {author} {\bibfnamefont {J.}~\bibnamefont {Hone}}, \bibinfo {author}
  {\bibfnamefont {H.~L.}\ \bibnamefont {Stormer}}, \ and\ \bibinfo {author}
  {\bibfnamefont {P.}~\bibnamefont {Kim}},\ }\href {\doibase
  10.1103/PhysRevLett.101.096802} {\bibfield  {journal} {\bibinfo  {journal}
  {Phys. Rev. Lett.}\ }\textbf {\bibinfo {volume} {101}},\ \bibinfo {pages}
  {096802} (\bibinfo {year} {2008})}\BibitemShut {NoStop}%
\bibitem [{\citenamefont {Du}\ \emph {et~al.}(2008)\citenamefont {Du},
  \citenamefont {Skachko}, \citenamefont {Barker},\ and\ \citenamefont
  {Andrei}}]{Du:NNano2008}%
  \BibitemOpen
  \bibfield  {author} {\bibinfo {author} {\bibfnamefont {X.}~\bibnamefont
  {Du}}, \bibinfo {author} {\bibfnamefont {I.}~\bibnamefont {Skachko}},
  \bibinfo {author} {\bibfnamefont {A.}~\bibnamefont {Barker}}, \ and\ \bibinfo
  {author} {\bibfnamefont {E.~Y.}\ \bibnamefont {Andrei}},\ }\href {\doibase
  10.1038/nnano.2008.199} {\bibfield  {journal} {\bibinfo  {journal} {Nature
  Nanotechnology}\ }\textbf {\bibinfo {volume} {3}},\ \bibinfo {pages} {491 }
  (\bibinfo {year} {2008})}\BibitemShut {NoStop}%
\bibitem [{\citenamefont {Lin}\ \emph {et~al.}(2009)\citenamefont {Lin},
  \citenamefont {Jenkins}, \citenamefont {Valdes-Garcia}, \citenamefont
  {Small}, \citenamefont {Farmer},\ and\ \citenamefont
  {Avouris}}]{Lin:NanoLet2009}%
  \BibitemOpen
  \bibfield  {author} {\bibinfo {author} {\bibfnamefont {Y.-M.}\ \bibnamefont
  {Lin}}, \bibinfo {author} {\bibfnamefont {K.~A.}\ \bibnamefont {Jenkins}},
  \bibinfo {author} {\bibfnamefont {A.}~\bibnamefont {Valdes-Garcia}}, \bibinfo
  {author} {\bibfnamefont {J.~P.}\ \bibnamefont {Small}}, \bibinfo {author}
  {\bibfnamefont {D.~B.}\ \bibnamefont {Farmer}}, \ and\ \bibinfo {author}
  {\bibfnamefont {P.}~\bibnamefont {Avouris}},\ }\href {\doibase
  10.1021/nl803316h} {\bibfield  {journal} {\bibinfo  {journal} {Nano Letters}\
  }\textbf {\bibinfo {volume} {9}},\ \bibinfo {pages} {422} (\bibinfo {year}
  {2009})}\BibitemShut {NoStop}%
\bibitem [{\citenamefont {Liao}\ \emph {et~al.}(2010)\citenamefont {Liao},
  \citenamefont {Lin}, \citenamefont {Bao}, \citenamefont {Cheng},
  \citenamefont {Bai}, \citenamefont {Liu}, \citenamefont {Qu}, \citenamefont
  {Wang}, \citenamefont {Huang},\ and\ \citenamefont {Duan}}]{Liao:Nature2010}%
  \BibitemOpen
  \bibfield  {author} {\bibinfo {author} {\bibfnamefont {L.}~\bibnamefont
  {Liao}}, \bibinfo {author} {\bibfnamefont {Y.-C.}\ \bibnamefont {Lin}},
  \bibinfo {author} {\bibfnamefont {M.}~\bibnamefont {Bao}}, \bibinfo {author}
  {\bibfnamefont {R.}~\bibnamefont {Cheng}}, \bibinfo {author} {\bibfnamefont
  {J.}~\bibnamefont {Bai}}, \bibinfo {author} {\bibfnamefont {Y.}~\bibnamefont
  {Liu}}, \bibinfo {author} {\bibfnamefont {Y.}~\bibnamefont {Qu}}, \bibinfo
  {author} {\bibfnamefont {K.~L.}\ \bibnamefont {Wang}}, \bibinfo {author}
  {\bibfnamefont {Y.}~\bibnamefont {Huang}}, \ and\ \bibinfo {author}
  {\bibfnamefont {X.}~\bibnamefont {Duan}},\ }\href {\doibase
  10.1038/nature09405} {\bibfield  {journal} {\bibinfo  {journal} {Nature}\
  }\textbf {\bibinfo {volume} {467}},\ \bibinfo {pages} {305} (\bibinfo {year}
  {2010})}\BibitemShut {NoStop}%
\bibitem [{\citenamefont {Lin}\ \emph {et~al.}(2011)\citenamefont {Lin},
  \citenamefont {Valdes-Garcia}, \citenamefont {Han}, \citenamefont {Farmer},
  \citenamefont {Meric}, \citenamefont {Sun}, \citenamefont {Wu}, \citenamefont
  {Dimitrakopoulos}, \citenamefont {Grill}, \citenamefont {Avouris},\ and\
  \citenamefont {Jenkins}}]{Lin:Science2011}%
  \BibitemOpen
  \bibfield  {author} {\bibinfo {author} {\bibfnamefont {Y.-M.}\ \bibnamefont
  {Lin}}, \bibinfo {author} {\bibfnamefont {A.}~\bibnamefont {Valdes-Garcia}},
  \bibinfo {author} {\bibfnamefont {S.-J.}\ \bibnamefont {Han}}, \bibinfo
  {author} {\bibfnamefont {D.~B.}\ \bibnamefont {Farmer}}, \bibinfo {author}
  {\bibfnamefont {I.}~\bibnamefont {Meric}}, \bibinfo {author} {\bibfnamefont
  {Y.}~\bibnamefont {Sun}}, \bibinfo {author} {\bibfnamefont {Y.}~\bibnamefont
  {Wu}}, \bibinfo {author} {\bibfnamefont {C.}~\bibnamefont {Dimitrakopoulos}},
  \bibinfo {author} {\bibfnamefont {A.}~\bibnamefont {Grill}}, \bibinfo
  {author} {\bibfnamefont {P.}~\bibnamefont {Avouris}}, \ and\ \bibinfo
  {author} {\bibfnamefont {K.~A.}\ \bibnamefont {Jenkins}},\ }\href {\doibase
  10.1126/science.1204428} {\bibfield  {journal} {\bibinfo  {journal}
  {Science}\ }\textbf {\bibinfo {volume} {332}},\ \bibinfo {pages} {1294}
  (\bibinfo {year} {2011})}\BibitemShut {NoStop}%
\bibitem [{\citenamefont {Schwierz}(2010)}]{Schwierz:NatNano2010}%
  \BibitemOpen
  \bibfield  {author} {\bibinfo {author} {\bibfnamefont {F.}~\bibnamefont
  {Schwierz}},\ }\href {\doibase 10.1038/nnano.2010.89} {\bibfield  {journal}
  {\bibinfo  {journal} {Nature Nanotechnology}\ }\textbf {\bibinfo {volume}
  {5}},\ \bibinfo {pages} {487} (\bibinfo {year} {2010})}\BibitemShut {NoStop}%
\bibitem [{\citenamefont {Adam}\ \emph {et~al.}(2009)\citenamefont {Adam},
  \citenamefont {Hwang}, \citenamefont {Rossi},\ and\ \citenamefont
  {Sarma}}]{adam:ssc2009}%
  \BibitemOpen
  \bibfield  {author} {\bibinfo {author} {\bibfnamefont {S.}~\bibnamefont
  {Adam}}, \bibinfo {author} {\bibfnamefont {E.}~\bibnamefont {Hwang}},
  \bibinfo {author} {\bibfnamefont {E.}~\bibnamefont {Rossi}}, \ and\ \bibinfo
  {author} {\bibfnamefont {S.~D.}\ \bibnamefont {Sarma}},\ }\href {\doibase
  http://dx.doi.org/10.1016/j.ssc.2009.02.041} {\bibfield  {journal} {\bibinfo
  {journal} {Solid State Communications}\ }\textbf {\bibinfo {volume} {149}},\
  \bibinfo {pages} {1072 } (\bibinfo {year} {2009})}\BibitemShut {NoStop}%
\bibitem [{\citenamefont {Ponomarenko}\ \emph {et~al.}(2009)\citenamefont
  {Ponomarenko}, \citenamefont {Yang}, \citenamefont {Mohiuddin}, \citenamefont
  {Katsnelson}, \citenamefont {Novoselov}, \citenamefont {Morozov},
  \citenamefont {Zhukov}, \citenamefont {Schedin}, \citenamefont {Hill},\ and\
  \citenamefont {Geim}}]{ponomarenko:prl2009}%
  \BibitemOpen
  \bibfield  {author} {\bibinfo {author} {\bibfnamefont {L.~A.}\ \bibnamefont
  {Ponomarenko}}, \bibinfo {author} {\bibfnamefont {R.}~\bibnamefont {Yang}},
  \bibinfo {author} {\bibfnamefont {T.~M.}\ \bibnamefont {Mohiuddin}}, \bibinfo
  {author} {\bibfnamefont {M.~I.}\ \bibnamefont {Katsnelson}}, \bibinfo
  {author} {\bibfnamefont {K.~S.}\ \bibnamefont {Novoselov}}, \bibinfo {author}
  {\bibfnamefont {S.~V.}\ \bibnamefont {Morozov}}, \bibinfo {author}
  {\bibfnamefont {A.~A.}\ \bibnamefont {Zhukov}}, \bibinfo {author}
  {\bibfnamefont {F.}~\bibnamefont {Schedin}}, \bibinfo {author} {\bibfnamefont
  {E.~W.}\ \bibnamefont {Hill}}, \ and\ \bibinfo {author} {\bibfnamefont
  {A.~K.}\ \bibnamefont {Geim}},\ }\href {\doibase
  10.1103/PhysRevLett.102.206603} {\bibfield  {journal} {\bibinfo  {journal}
  {Phys. Rev. Lett.}\ }\textbf {\bibinfo {volume} {102}},\ \bibinfo {pages}
  {206603} (\bibinfo {year} {2009})}\BibitemShut {NoStop}%
\bibitem [{\citenamefont {Katsnelson}\ and\ \citenamefont
  {Geim}(2008)}]{katsnelson:pht08}%
  \BibitemOpen
  \bibfield  {author} {\bibinfo {author} {\bibfnamefont {M.}~\bibnamefont
  {Katsnelson}}\ and\ \bibinfo {author} {\bibfnamefont {A.}~\bibnamefont
  {Geim}},\ }\href {\doibase 10.1098/rsta.2007.2157} {\bibfield  {journal}
  {\bibinfo  {journal} {Philos. Trans. R. Soc. A}\ }\textbf {\bibinfo {volume}
  {366}},\ \bibinfo {pages} {195} (\bibinfo {year} {2008})}\BibitemShut
  {NoStop}%
\bibitem [{\citenamefont {Meyer}\ \emph {et~al.}(2007)\citenamefont {Meyer},
  \citenamefont {Geim}, \citenamefont {Katsnelson}, \citenamefont {Novoselov},
  \citenamefont {Booth},\ and\ \citenamefont {Roth}}]{Meyer:nat07}%
  \BibitemOpen
  \bibfield  {author} {\bibinfo {author} {\bibfnamefont {J.~C.}\ \bibnamefont
  {Meyer}}, \bibinfo {author} {\bibfnamefont {A.~K.}\ \bibnamefont {Geim}},
  \bibinfo {author} {\bibfnamefont {M.~I.}\ \bibnamefont {Katsnelson}},
  \bibinfo {author} {\bibfnamefont {K.~S.}\ \bibnamefont {Novoselov}}, \bibinfo
  {author} {\bibfnamefont {T.~J.}\ \bibnamefont {Booth}}, \ and\ \bibinfo
  {author} {\bibfnamefont {S.}~\bibnamefont {Roth}},\ }\href {\doibase
  10.1038/nature05545} {\bibfield  {journal} {\bibinfo  {journal} {Nature}\
  }\textbf {\bibinfo {volume} {446}},\ \bibinfo {pages} {60} (\bibinfo {year}
  {2007})}\BibitemShut {NoStop}%
\bibitem [{\citenamefont {Stolyarova}\ \emph {et~al.}(2007)\citenamefont
  {Stolyarova}, \citenamefont {Rim}, \citenamefont {Ryu}, \citenamefont
  {Maultzsch}, \citenamefont {Kim}, \citenamefont {Brus}, \citenamefont
  {Heinz}, \citenamefont {Hybertsen},\ and\ \citenamefont
  {Flynn}}]{Stolyarova:pnas07}%
  \BibitemOpen
  \bibfield  {author} {\bibinfo {author} {\bibfnamefont {E.}~\bibnamefont
  {Stolyarova}}, \bibinfo {author} {\bibfnamefont {K.~T.}\ \bibnamefont {Rim}},
  \bibinfo {author} {\bibfnamefont {S.}~\bibnamefont {Ryu}}, \bibinfo {author}
  {\bibfnamefont {J.}~\bibnamefont {Maultzsch}}, \bibinfo {author}
  {\bibfnamefont {P.}~\bibnamefont {Kim}}, \bibinfo {author} {\bibfnamefont
  {L.~E.}\ \bibnamefont {Brus}}, \bibinfo {author} {\bibfnamefont {T.~F.}\
  \bibnamefont {Heinz}}, \bibinfo {author} {\bibfnamefont {M.~S.}\ \bibnamefont
  {Hybertsen}}, \ and\ \bibinfo {author} {\bibfnamefont {G.~W.}\ \bibnamefont
  {Flynn}},\ }\href {\doibase 10.1073/pnas.0703337104} {\bibfield  {journal}
  {\bibinfo  {journal} {Proc. Natl. Acad. Sci. U.S.A.}\ }\textbf {\bibinfo
  {volume} {104}},\ \bibinfo {pages} {9209} (\bibinfo {year}
  {2007})}\BibitemShut {NoStop}%
\bibitem [{\citenamefont {Ishigami}\ \emph {et~al.}(2007)\citenamefont
  {Ishigami}, \citenamefont {Chen}, \citenamefont {Cullen}, \citenamefont
  {Fuhrer},\ and\ \citenamefont {Williams}}]{ishigami:nanolet07}%
  \BibitemOpen
  \bibfield  {author} {\bibinfo {author} {\bibfnamefont {M.}~\bibnamefont
  {Ishigami}}, \bibinfo {author} {\bibfnamefont {J.~H.}\ \bibnamefont {Chen}},
  \bibinfo {author} {\bibfnamefont {W.~G.}\ \bibnamefont {Cullen}}, \bibinfo
  {author} {\bibfnamefont {M.~S.}\ \bibnamefont {Fuhrer}}, \ and\ \bibinfo
  {author} {\bibfnamefont {E.~D.}\ \bibnamefont {Williams}},\ }\href {\doibase
  10.1021/nl070613a} {\bibfield  {journal} {\bibinfo  {journal} {Nano Letters}\
  }\textbf {\bibinfo {volume} {7}},\ \bibinfo {pages} {1643} (\bibinfo {year}
  {2007})}\BibitemShut {NoStop}%
\bibitem [{\citenamefont {Geringer}\ \emph {et~al.}(2009)\citenamefont
  {Geringer}, \citenamefont {Liebmann}, \citenamefont {Echtermeyer},
  \citenamefont {Runte}, \citenamefont {Schmidt}, \citenamefont {R\"uckamp},
  \citenamefont {Lemme},\ and\ \citenamefont {Morgenstern}}]{geringer:prl09}%
  \BibitemOpen
  \bibfield  {author} {\bibinfo {author} {\bibfnamefont {V.}~\bibnamefont
  {Geringer}}, \bibinfo {author} {\bibfnamefont {M.}~\bibnamefont {Liebmann}},
  \bibinfo {author} {\bibfnamefont {T.}~\bibnamefont {Echtermeyer}}, \bibinfo
  {author} {\bibfnamefont {S.}~\bibnamefont {Runte}}, \bibinfo {author}
  {\bibfnamefont {M.}~\bibnamefont {Schmidt}}, \bibinfo {author} {\bibfnamefont
  {R.}~\bibnamefont {R\"uckamp}}, \bibinfo {author} {\bibfnamefont {M.~C.}\
  \bibnamefont {Lemme}}, \ and\ \bibinfo {author} {\bibfnamefont
  {M.}~\bibnamefont {Morgenstern}},\ }\href {\doibase
  10.1103/PhysRevLett.102.076102} {\bibfield  {journal} {\bibinfo  {journal}
  {Phys. Rev. Lett.}\ }\textbf {\bibinfo {volume} {102}},\ \bibinfo {pages}
  {076102} (\bibinfo {year} {2009})}\BibitemShut {NoStop}%
\bibitem [{\citenamefont {Cullen}\ \emph {et~al.}(2010)\citenamefont {Cullen},
  \citenamefont {Yamamoto}, \citenamefont {Burson}, \citenamefont {Chen},
  \citenamefont {Jang}, \citenamefont {Li}, \citenamefont {Fuhrer},\ and\
  \citenamefont {Williams}}]{Cullen:prl2010}%
  \BibitemOpen
  \bibfield  {author} {\bibinfo {author} {\bibfnamefont {W.~G.}\ \bibnamefont
  {Cullen}}, \bibinfo {author} {\bibfnamefont {M.}~\bibnamefont {Yamamoto}},
  \bibinfo {author} {\bibfnamefont {K.~M.}\ \bibnamefont {Burson}}, \bibinfo
  {author} {\bibfnamefont {J.~H.}\ \bibnamefont {Chen}}, \bibinfo {author}
  {\bibfnamefont {C.}~\bibnamefont {Jang}}, \bibinfo {author} {\bibfnamefont
  {L.}~\bibnamefont {Li}}, \bibinfo {author} {\bibfnamefont {M.~S.}\
  \bibnamefont {Fuhrer}}, \ and\ \bibinfo {author} {\bibfnamefont {E.~D.}\
  \bibnamefont {Williams}},\ }\href {\doibase 10.1103/PhysRevLett.105.215504}
  {\bibfield  {journal} {\bibinfo  {journal} {Phys. Rev. Lett.}\ }\textbf
  {\bibinfo {volume} {105}},\ \bibinfo {pages} {215504} (\bibinfo {year}
  {2010})}\BibitemShut {NoStop}%
\bibitem [{\citenamefont {A.~Fasolino}\ and\ \citenamefont
  {Katsnelson}(2007)}]{fasolino:natm07}%
  \BibitemOpen
  \bibfield  {author} {\bibinfo {author} {\bibfnamefont {J.~H.~L.}\
  \bibnamefont {A.~Fasolino}}\ and\ \bibinfo {author} {\bibfnamefont {M.~I.}\
  \bibnamefont {Katsnelson}},\ }\href {\doibase
  http://dx.doi.org/10.1038/nmat2011} {\bibfield  {journal} {\bibinfo
  {journal} {Nature Materials}\ }\textbf {\bibinfo {volume} {6}},\ \bibinfo
  {pages} {858 } (\bibinfo {year} {2007})}\BibitemShut {NoStop}%
\bibitem [{\citenamefont {Abedpour}\ \emph {et~al.}(2007)\citenamefont
  {Abedpour}, \citenamefont {Neek-Amal}, \citenamefont {Asgari}, \citenamefont
  {Shahbazi}, \citenamefont {Nafari},\ and\ \citenamefont
  {Tabar}}]{Abdepour:prb2007}%
  \BibitemOpen
  \bibfield  {author} {\bibinfo {author} {\bibfnamefont {N.}~\bibnamefont
  {Abedpour}}, \bibinfo {author} {\bibfnamefont {M.}~\bibnamefont {Neek-Amal}},
  \bibinfo {author} {\bibfnamefont {R.}~\bibnamefont {Asgari}}, \bibinfo
  {author} {\bibfnamefont {F.}~\bibnamefont {Shahbazi}}, \bibinfo {author}
  {\bibfnamefont {N.}~\bibnamefont {Nafari}}, \ and\ \bibinfo {author}
  {\bibfnamefont {M.~R.~R.}\ \bibnamefont {Tabar}},\ }\href {\doibase
  10.1103/PhysRevB.76.195407} {\bibfield  {journal} {\bibinfo  {journal} {Phys.
  Rev. B}\ }\textbf {\bibinfo {volume} {76}},\ \bibinfo {pages} {195407}
  (\bibinfo {year} {2007})}\BibitemShut {NoStop}%
\bibitem [{\citenamefont {Kim}\ and\ \citenamefont {Neto}(2008)}]{Kim:EPL2008}%
  \BibitemOpen
  \bibfield  {author} {\bibinfo {author} {\bibfnamefont {E.-A.}\ \bibnamefont
  {Kim}}\ and\ \bibinfo {author} {\bibfnamefont {A.~H.~C.}\ \bibnamefont
  {Neto}},\ }\href {http://stacks.iop.org/0295-5075/84/i=5/a=57007} {\bibfield
  {journal} {\bibinfo  {journal} {Europhys. Lett.}\ }\textbf {\bibinfo {volume}
  {84}},\ \bibinfo {pages} {57007} (\bibinfo {year} {2008})}\BibitemShut
  {NoStop}%
\bibitem [{\citenamefont {Gazit}(2009{\natexlab{a}})}]{gazit:prb2009}%
  \BibitemOpen
  \bibfield  {author} {\bibinfo {author} {\bibfnamefont {D.}~\bibnamefont
  {Gazit}},\ }\href {\doibase 10.1103/PhysRevB.79.113411} {\bibfield  {journal}
  {\bibinfo  {journal} {Phys. Rev. B}\ }\textbf {\bibinfo {volume} {79}},\
  \bibinfo {pages} {113411} (\bibinfo {year} {2009}{\natexlab{a}})}\BibitemShut
  {NoStop}%
\bibitem [{\citenamefont {Los}\ \emph {et~al.}(2009)\citenamefont {Los},
  \citenamefont {Katsnelson}, \citenamefont {Yazyev}, \citenamefont
  {Zakharchenko},\ and\ \citenamefont {Fasolino}}]{Los:prb2009}%
  \BibitemOpen
  \bibfield  {author} {\bibinfo {author} {\bibfnamefont {J.~H.}\ \bibnamefont
  {Los}}, \bibinfo {author} {\bibfnamefont {M.~I.}\ \bibnamefont {Katsnelson}},
  \bibinfo {author} {\bibfnamefont {O.~V.}\ \bibnamefont {Yazyev}}, \bibinfo
  {author} {\bibfnamefont {K.~V.}\ \bibnamefont {Zakharchenko}}, \ and\
  \bibinfo {author} {\bibfnamefont {A.}~\bibnamefont {Fasolino}},\ }\href
  {\doibase 10.1103/PhysRevB.80.121405} {\bibfield  {journal} {\bibinfo
  {journal} {Phys. Rev. B}\ }\textbf {\bibinfo {volume} {80}},\ \bibinfo
  {pages} {121405} (\bibinfo {year} {2009})}\BibitemShut {NoStop}%
\bibitem [{Note1()}]{Note1}%
  \BibitemOpen
  \bibinfo {note} {The height correlation function typically scales like $\left
  < (h(\protect \mathbf {r})-h(0))^2\right >\sim 2\left <h^2\right >\left ( 1 -
  e^{(r/\xi )^{2H}} \right )$}\BibitemShut {NoStop}%
\bibitem [{\citenamefont {Guinea}\ \emph
  {et~al.}(2008{\natexlab{a}})\citenamefont {Guinea}, \citenamefont
  {Horovitz},\ and\ \citenamefont {Le~Doussal}}]{guinea:prb08a}%
  \BibitemOpen
  \bibfield  {author} {\bibinfo {author} {\bibfnamefont {F.}~\bibnamefont
  {Guinea}}, \bibinfo {author} {\bibfnamefont {B.}~\bibnamefont {Horovitz}}, \
  and\ \bibinfo {author} {\bibfnamefont {P.}~\bibnamefont {Le~Doussal}},\
  }\href {\doibase 10.1103/PhysRevB.77.205421} {\bibfield  {journal} {\bibinfo
  {journal} {Phys. Rev. B}\ }\textbf {\bibinfo {volume} {77}},\ \bibinfo
  {pages} {205421} (\bibinfo {year} {2008}{\natexlab{a}})}\BibitemShut
  {NoStop}%
\bibitem [{\citenamefont {Guinea}\ \emph
  {et~al.}(2008{\natexlab{b}})\citenamefont {Guinea}, \citenamefont
  {Katsnelson},\ and\ \citenamefont {Vozmediano}}]{guinea:prb08b}%
  \BibitemOpen
  \bibfield  {author} {\bibinfo {author} {\bibfnamefont {F.}~\bibnamefont
  {Guinea}}, \bibinfo {author} {\bibfnamefont {M.~I.}\ \bibnamefont
  {Katsnelson}}, \ and\ \bibinfo {author} {\bibfnamefont {M.~A.~H.}\
  \bibnamefont {Vozmediano}},\ }\href {\doibase 10.1103/PhysRevB.77.075422}
  {\bibfield  {journal} {\bibinfo  {journal} {Phys. Rev. B}\ }\textbf {\bibinfo
  {volume} {77}},\ \bibinfo {pages} {075422} (\bibinfo {year}
  {2008}{\natexlab{b}})}\BibitemShut {NoStop}%
\bibitem [{\citenamefont {Vozmediano}\ \emph {et~al.}(2010)\citenamefont
  {Vozmediano}, \citenamefont {Katsnelson},\ and\ \citenamefont
  {Guinea}}]{vozmediano:physrep2010}%
  \BibitemOpen
  \bibfield  {author} {\bibinfo {author} {\bibfnamefont {M.}~\bibnamefont
  {Vozmediano}}, \bibinfo {author} {\bibfnamefont {M.}~\bibnamefont
  {Katsnelson}}, \ and\ \bibinfo {author} {\bibfnamefont {F.}~\bibnamefont
  {Guinea}},\ }\href {\doibase http://dx.doi.org/10.1016/j.physrep.2010.07.003}
  {\bibfield  {journal} {\bibinfo  {journal} {Physics Reports}\ }\textbf
  {\bibinfo {volume} {496}},\ \bibinfo {pages} {109 } (\bibinfo {year}
  {2010})}\BibitemShut {NoStop}%
\bibitem [{\citenamefont {Morozov}\ \emph {et~al.}(2006)\citenamefont
  {Morozov}, \citenamefont {Novoselov}, \citenamefont {Katsnelson},
  \citenamefont {Schedin}, \citenamefont {Ponomarenko}, \citenamefont {Jiang},\
  and\ \citenamefont {Geim}}]{Morozov:prl06}%
  \BibitemOpen
  \bibfield  {author} {\bibinfo {author} {\bibfnamefont {S.~V.}\ \bibnamefont
  {Morozov}}, \bibinfo {author} {\bibfnamefont {K.~S.}\ \bibnamefont
  {Novoselov}}, \bibinfo {author} {\bibfnamefont {M.~I.}\ \bibnamefont
  {Katsnelson}}, \bibinfo {author} {\bibfnamefont {F.}~\bibnamefont {Schedin}},
  \bibinfo {author} {\bibfnamefont {L.~A.}\ \bibnamefont {Ponomarenko}},
  \bibinfo {author} {\bibfnamefont {D.}~\bibnamefont {Jiang}}, \ and\ \bibinfo
  {author} {\bibfnamefont {A.~K.}\ \bibnamefont {Geim}},\ }\href {\doibase
  10.1103/PhysRevLett.97.016801} {\bibfield  {journal} {\bibinfo  {journal}
  {Phys. Rev. Lett.}\ }\textbf {\bibinfo {volume} {97}},\ \bibinfo {pages}
  {016801} (\bibinfo {year} {2006})}\BibitemShut {NoStop}%
\bibitem [{\citenamefont {Gibertini}\ \emph {et~al.}(2010)\citenamefont
  {Gibertini}, \citenamefont {Tomadin}, \citenamefont {Polini}, \citenamefont
  {Fasolino},\ and\ \citenamefont {Katsnelson}}]{gibertini:prb10}%
  \BibitemOpen
  \bibfield  {author} {\bibinfo {author} {\bibfnamefont {M.}~\bibnamefont
  {Gibertini}}, \bibinfo {author} {\bibfnamefont {A.}~\bibnamefont {Tomadin}},
  \bibinfo {author} {\bibfnamefont {M.}~\bibnamefont {Polini}}, \bibinfo
  {author} {\bibfnamefont {A.}~\bibnamefont {Fasolino}}, \ and\ \bibinfo
  {author} {\bibfnamefont {M.~I.}\ \bibnamefont {Katsnelson}},\ }\href
  {\doibase 10.1103/PhysRevB.81.125437} {\bibfield  {journal} {\bibinfo
  {journal} {Phys. Rev. B}\ }\textbf {\bibinfo {volume} {81}},\ \bibinfo
  {pages} {125437} (\bibinfo {year} {2010})}\BibitemShut {NoStop}%
\bibitem [{Note2()}]{Note2}%
  \BibitemOpen
  \bibinfo {note} {Longer length scales were obtained by the same group\cite
  {gibertini:prb(r)2012} when their formalism was applied to the experimental
  structure of Ref.~\protect \rev@citealpnum {geringer:prl09}.}\BibitemShut
  {Stop}%
\bibitem [{\citenamefont {Partovi-Azar}\ \emph {et~al.}(2011)\citenamefont
  {Partovi-Azar}, \citenamefont {Nafari},\ and\ \citenamefont
  {Tabar}}]{partovi:prb2011}%
  \BibitemOpen
  \bibfield  {author} {\bibinfo {author} {\bibfnamefont {P.}~\bibnamefont
  {Partovi-Azar}}, \bibinfo {author} {\bibfnamefont {N.}~\bibnamefont
  {Nafari}}, \ and\ \bibinfo {author} {\bibfnamefont {M.~R.~R.}\ \bibnamefont
  {Tabar}},\ }\href {\doibase 10.1103/PhysRevB.83.165434} {\bibfield  {journal}
  {\bibinfo  {journal} {Phys. Rev. B}\ }\textbf {\bibinfo {volume} {83}},\
  \bibinfo {pages} {165434} (\bibinfo {year} {2011})}\BibitemShut {NoStop}%
\bibitem [{\citenamefont {Gazit}(2009{\natexlab{b}})}]{gazit:prb(R)2009}%
  \BibitemOpen
  \bibfield  {author} {\bibinfo {author} {\bibfnamefont {D.}~\bibnamefont
  {Gazit}},\ }\href {\doibase 10.1103/PhysRevB.80.161406} {\bibfield  {journal}
  {\bibinfo  {journal} {Phys. Rev. B}\ }\textbf {\bibinfo {volume} {80}},\
  \bibinfo {pages} {161406} (\bibinfo {year} {2009}{\natexlab{b}})}\BibitemShut
  {NoStop}%
\bibitem [{\citenamefont {Deshpande}\ \emph {et~al.}(2009)\citenamefont
  {Deshpande}, \citenamefont {Bao}, \citenamefont {Miao}, \citenamefont {Lau},\
  and\ \citenamefont {LeRoy}}]{desphande:prb2009}%
  \BibitemOpen
  \bibfield  {author} {\bibinfo {author} {\bibfnamefont {A.}~\bibnamefont
  {Deshpande}}, \bibinfo {author} {\bibfnamefont {W.}~\bibnamefont {Bao}},
  \bibinfo {author} {\bibfnamefont {F.}~\bibnamefont {Miao}}, \bibinfo {author}
  {\bibfnamefont {C.~N.}\ \bibnamefont {Lau}}, \ and\ \bibinfo {author}
  {\bibfnamefont {B.~J.}\ \bibnamefont {LeRoy}},\ }\href {\doibase
  10.1103/PhysRevB.79.205411} {\bibfield  {journal} {\bibinfo  {journal} {Phys.
  Rev. B}\ }\textbf {\bibinfo {volume} {79}},\ \bibinfo {pages} {205411}
  (\bibinfo {year} {2009})}\BibitemShut {NoStop}%
\bibitem [{\citenamefont {Zhang}\ \emph {et~al.}(2009)\citenamefont {Zhang},
  \citenamefont {Brar}, \citenamefont {Girit}, \citenamefont {Zettl},\ and\
  \citenamefont {Crommie}}]{zhang:natphys2009}%
  \BibitemOpen
  \bibfield  {author} {\bibinfo {author} {\bibfnamefont {Y.}~\bibnamefont
  {Zhang}}, \bibinfo {author} {\bibfnamefont {V.~W.}\ \bibnamefont {Brar}},
  \bibinfo {author} {\bibfnamefont {C.}~\bibnamefont {Girit}}, \bibinfo
  {author} {\bibfnamefont {A.}~\bibnamefont {Zettl}}, \ and\ \bibinfo {author}
  {\bibfnamefont {M.~F.}\ \bibnamefont {Crommie}},\ }\href {\doibase
  10.1038/nphys1365} {\bibfield  {journal} {\bibinfo  {journal} {Nature
  Physics}\ }\textbf {\bibinfo {volume} {5}},\ \bibinfo {pages} {722} (\bibinfo
  {year} {2009})}\BibitemShut {NoStop}%
\bibitem [{\citenamefont {Isacsson}\ \emph {et~al.}(2008)\citenamefont
  {Isacsson}, \citenamefont {Jonsson}, \citenamefont {Kinaret},\ and\
  \citenamefont {Jonson}}]{Isacsson:prb2008}%
  \BibitemOpen
  \bibfield  {author} {\bibinfo {author} {\bibfnamefont {A.}~\bibnamefont
  {Isacsson}}, \bibinfo {author} {\bibfnamefont {L.~M.}\ \bibnamefont
  {Jonsson}}, \bibinfo {author} {\bibfnamefont {J.~M.}\ \bibnamefont
  {Kinaret}}, \ and\ \bibinfo {author} {\bibfnamefont {M.}~\bibnamefont
  {Jonson}},\ }\href {\doibase 10.1103/PhysRevB.77.035423} {\bibfield
  {journal} {\bibinfo  {journal} {Phys. Rev. B}\ }\textbf {\bibinfo {volume}
  {77}},\ \bibinfo {pages} {035423} (\bibinfo {year} {2008})}\BibitemShut
  {NoStop}%
\bibitem [{\citenamefont {Zwierzycki}(2012)}]{zwierzycki:app2012}%
  \BibitemOpen
  \bibfield  {author} {\bibinfo {author} {\bibfnamefont {M.}~\bibnamefont
  {Zwierzycki}},\ }\href
  {http://przyrbwn.icm.edu.pl/APP/ABSTR/121/a121-5-79.html} {\bibfield
  {journal} {\bibinfo  {journal} {Acta. Phys. Pol. A}\ }\textbf {\bibinfo
  {volume} {121}},\ \bibinfo {pages} {1246 } (\bibinfo {year}
  {2012})}\BibitemShut {NoStop}%
\bibitem [{\citenamefont {Klos}\ \emph {et~al.}(2009)\citenamefont {Klos},
  \citenamefont {Shylau}, \citenamefont {Zozoulenko}, \citenamefont {Xu},\ and\
  \citenamefont {Heinzel}}]{Klos:PRB2009}%
  \BibitemOpen
  \bibfield  {author} {\bibinfo {author} {\bibfnamefont {J.~W.}\ \bibnamefont
  {Klos}}, \bibinfo {author} {\bibfnamefont {A.~A.}\ \bibnamefont {Shylau}},
  \bibinfo {author} {\bibfnamefont {I.~V.}\ \bibnamefont {Zozoulenko}},
  \bibinfo {author} {\bibfnamefont {H.}~\bibnamefont {Xu}}, \ and\ \bibinfo
  {author} {\bibfnamefont {T.}~\bibnamefont {Heinzel}},\ }\href {\doibase
  10.1103/PhysRevB.80.245432} {\bibfield  {journal} {\bibinfo  {journal} {Phys.
  Rev. B}\ }\textbf {\bibinfo {volume} {80}},\ \bibinfo {pages} {245432}
  (\bibinfo {year} {2009})}\BibitemShut {NoStop}%
\bibitem [{\citenamefont {Zhu}\ and\ \citenamefont
  {Lv}(2013)}]{Zhu:PhysLettA2013}%
  \BibitemOpen
  \bibfield  {author} {\bibinfo {author} {\bibfnamefont {W.}~\bibnamefont
  {Zhu}}\ and\ \bibinfo {author} {\bibfnamefont {B.}~\bibnamefont {Lv}},\
  }\href {\doibase http://dx.doi.org/10.1016/j.physleta.2013.04.048} {\bibfield
   {journal} {\bibinfo  {journal} {Physics Letters A}\ }\textbf {\bibinfo
  {volume} {377}},\ \bibinfo {pages} {1649 } (\bibinfo {year}
  {2013})}\BibitemShut {NoStop}%
\bibitem [{\citenamefont {Xia}\ \emph {et~al.}(2006)\citenamefont {Xia},
  \citenamefont {Zwierzycki}, \citenamefont {Talanana}, \citenamefont {Kelly},\
  and\ \citenamefont {Bauer}}]{Xia:prb06}%
  \BibitemOpen
  \bibfield  {author} {\bibinfo {author} {\bibfnamefont {K.}~\bibnamefont
  {Xia}}, \bibinfo {author} {\bibfnamefont {M.}~\bibnamefont {Zwierzycki}},
  \bibinfo {author} {\bibfnamefont {M.}~\bibnamefont {Talanana}}, \bibinfo
  {author} {\bibfnamefont {P.~J.}\ \bibnamefont {Kelly}}, \ and\ \bibinfo
  {author} {\bibfnamefont {G.~E.~W.}\ \bibnamefont {Bauer}},\ }\href {\doibase
  10.1103/PhysRevB.73.064420} {\bibfield  {journal} {\bibinfo  {journal} {Phys.
  Rev. B}\ }\textbf {\bibinfo {volume} {73}},\ \bibinfo {pages} {064420}
  (\bibinfo {year} {2006})}\BibitemShut {NoStop}%
\bibitem [{\citenamefont {Zwierzycki}\ \emph {et~al.}(2008)\citenamefont
  {Zwierzycki}, \citenamefont {Khomyakov}, \citenamefont {Starikov},
  \citenamefont {Xia}, \citenamefont {Talanana}, \citenamefont {Xu},
  \citenamefont {Karpan}, \citenamefont {Marushchenko}, \citenamefont {Turek},
  \citenamefont {Bauer}, \citenamefont {Brocks},\ and\ \citenamefont
  {Kelly}}]{Zwierzycki:pssb08}%
  \BibitemOpen
  \bibfield  {author} {\bibinfo {author} {\bibfnamefont {M.}~\bibnamefont
  {Zwierzycki}}, \bibinfo {author} {\bibfnamefont {P.~A.}\ \bibnamefont
  {Khomyakov}}, \bibinfo {author} {\bibfnamefont {A.~A.}\ \bibnamefont
  {Starikov}}, \bibinfo {author} {\bibfnamefont {K.}~\bibnamefont {Xia}},
  \bibinfo {author} {\bibfnamefont {M.}~\bibnamefont {Talanana}}, \bibinfo
  {author} {\bibfnamefont {P.~X.}\ \bibnamefont {Xu}}, \bibinfo {author}
  {\bibfnamefont {V.~M.}\ \bibnamefont {Karpan}}, \bibinfo {author}
  {\bibfnamefont {I.}~\bibnamefont {Marushchenko}}, \bibinfo {author}
  {\bibfnamefont {I.}~\bibnamefont {Turek}}, \bibinfo {author} {\bibfnamefont
  {G.~E.~W.}\ \bibnamefont {Bauer}}, \bibinfo {author} {\bibfnamefont
  {G.}~\bibnamefont {Brocks}}, \ and\ \bibinfo {author} {\bibfnamefont {P.~J.}\
  \bibnamefont {Kelly}},\ }\href {\doibase 10.1002/pssb.200743359} {\bibfield
  {journal} {\bibinfo  {journal} {Phys. Stat. Sol. B}\ }\textbf {\bibinfo
  {volume} {245}},\ \bibinfo {pages} {623} (\bibinfo {year}
  {2008})}\BibitemShut {NoStop}%
\bibitem [{\citenamefont {Fournier}\ \emph {et~al.}(1982)\citenamefont
  {Fournier}, \citenamefont {Fussell},\ and\ \citenamefont
  {Carpenter}}]{Fournier:CommACM1982}%
  \BibitemOpen
  \bibfield  {author} {\bibinfo {author} {\bibfnamefont {A.}~\bibnamefont
  {Fournier}}, \bibinfo {author} {\bibfnamefont {D.}~\bibnamefont {Fussell}}, \
  and\ \bibinfo {author} {\bibfnamefont {L.}~\bibnamefont {Carpenter}},\ }\href
  {\doibase 10.1145/358523.358553} {\bibfield  {journal} {\bibinfo  {journal}
  {Commun. ACM}\ }\textbf {\bibinfo {volume} {25}},\ \bibinfo {pages} {371}
  (\bibinfo {year} {1982})}\BibitemShut {NoStop}%
\bibitem [{\citenamefont {Gibertini}\ \emph {et~al.}(2012)\citenamefont
  {Gibertini}, \citenamefont {Tomadin}, \citenamefont {Guinea}, \citenamefont
  {Katsnelson},\ and\ \citenamefont {Polini}}]{gibertini:prb(r)2012}%
  \BibitemOpen
  \bibfield  {author} {\bibinfo {author} {\bibfnamefont {M.}~\bibnamefont
  {Gibertini}}, \bibinfo {author} {\bibfnamefont {A.}~\bibnamefont {Tomadin}},
  \bibinfo {author} {\bibfnamefont {F.}~\bibnamefont {Guinea}}, \bibinfo
  {author} {\bibfnamefont {M.~I.}\ \bibnamefont {Katsnelson}}, \ and\ \bibinfo
  {author} {\bibfnamefont {M.}~\bibnamefont {Polini}},\ }\href {\doibase
  10.1103/PhysRevB.85.201405} {\bibfield  {journal} {\bibinfo  {journal} {Phys.
  Rev. B}\ }\textbf {\bibinfo {volume} {85}},\ \bibinfo {pages} {201405}
  (\bibinfo {year} {2012})}\BibitemShut {NoStop}%
\bibitem [{Note3()}]{Note3}%
  \BibitemOpen
  \bibinfo {note} {The value of $V_{G}$ is calculated assuming that rippling
  does not change the position of neutrality point.}\BibitemShut {Stop}%
\bibitem [{\citenamefont {Tworzyd\l{}o}\ \emph {et~al.}(2006)\citenamefont
  {Tworzyd\l{}o}, \citenamefont {Trauzettel}, \citenamefont {Titov},
  \citenamefont {Rycerz},\ and\ \citenamefont {Beenakker}}]{Tworzydlo:prl2006}%
  \BibitemOpen
  \bibfield  {author} {\bibinfo {author} {\bibfnamefont {J.}~\bibnamefont
  {Tworzyd\l{}o}}, \bibinfo {author} {\bibfnamefont {B.}~\bibnamefont
  {Trauzettel}}, \bibinfo {author} {\bibfnamefont {M.}~\bibnamefont {Titov}},
  \bibinfo {author} {\bibfnamefont {A.}~\bibnamefont {Rycerz}}, \ and\ \bibinfo
  {author} {\bibfnamefont {C.~W.~J.}\ \bibnamefont {Beenakker}},\ }\href
  {\doibase 10.1103/PhysRevLett.96.246802} {\bibfield  {journal} {\bibinfo
  {journal} {Phys. Rev. Lett.}\ }\textbf {\bibinfo {volume} {96}},\ \bibinfo
  {pages} {246802} (\bibinfo {year} {2006})}\BibitemShut {NoStop}%
\bibitem [{\citenamefont {Sharvin}(1965)}]{Sharvin:zetf65}%
  \BibitemOpen
  \bibfield  {author} {\bibinfo {author} {\bibfnamefont {Y.~V.}\ \bibnamefont
  {Sharvin}},\ }\href@noop {} {\bibfield  {journal} {\bibinfo  {journal} {Zh.
  Eksp. Teor. Fiz.}\ }\textbf {\bibinfo {volume} {48}},\ \bibinfo {pages} {984}
  (\bibinfo {year} {1965})},\ \bibinfo {note} {[Sov. Phys. JETP \textbf{21},
  655 (1965)]}\BibitemShut {NoStop}%
\bibitem [{\citenamefont {Datta}(1995)}]{Datta:95}%
  \BibitemOpen
  \bibfield  {author} {\bibinfo {author} {\bibfnamefont {S.}~\bibnamefont
  {Datta}},\ }\href@noop {} {\emph {\bibinfo {title} {Electronic Transport in
  Mesoscopic Systems}}}\ (\bibinfo  {publisher} {Cambridge University Press},\
  \bibinfo {address} {Cambridge},\ \bibinfo {year} {1995})\BibitemShut
  {NoStop}%
\bibitem [{\citenamefont {Xu}\ and\ \citenamefont {Xia}(2006)}]{xu:prb2006}%
  \BibitemOpen
  \bibfield  {author} {\bibinfo {author} {\bibfnamefont {P.~X.}\ \bibnamefont
  {Xu}}\ and\ \bibinfo {author} {\bibfnamefont {K.}~\bibnamefont {Xia}},\
  }\href {\doibase 10.1103/PhysRevB.74.184206} {\bibfield  {journal} {\bibinfo
  {journal} {Phys. Rev. B}\ }\textbf {\bibinfo {volume} {74}},\ \bibinfo
  {pages} {184206} (\bibinfo {year} {2006})}\BibitemShut {NoStop}%
\bibitem [{\citenamefont {Wehling}\ \emph {et~al.}(2008)\citenamefont
  {Wehling}, \citenamefont {Balatsky}, \citenamefont {Tsvelik}, \citenamefont
  {Katsnelson},\ and\ \citenamefont {Lichtenstein}}]{wehling:epl08}%
  \BibitemOpen
  \bibfield  {author} {\bibinfo {author} {\bibfnamefont {T.~O.}\ \bibnamefont
  {Wehling}}, \bibinfo {author} {\bibfnamefont {A.~V.}\ \bibnamefont
  {Balatsky}}, \bibinfo {author} {\bibfnamefont {A.~M.}\ \bibnamefont
  {Tsvelik}}, \bibinfo {author} {\bibfnamefont {M.~I.}\ \bibnamefont
  {Katsnelson}}, \ and\ \bibinfo {author} {\bibfnamefont {A.~I.}\ \bibnamefont
  {Lichtenstein}},\ }\href {\doibase 10.1209/0295-5075/84/17003} {\bibfield
  {journal} {\bibinfo  {journal} {Europhys. Lett.}\ }\textbf {\bibinfo {volume}
  {84}},\ \bibinfo {pages} {17003} (\bibinfo {year} {2008})}\BibitemShut
  {NoStop}%
\bibitem [{\citenamefont {Andersen}\ \emph {et~al.}(1985)\citenamefont
  {Andersen}, \citenamefont {Jepsen},\ and\ \citenamefont
  {Gl{\"{o}}tzel}}]{Andersen:85}%
  \BibitemOpen
  \bibfield  {author} {\bibinfo {author} {\bibfnamefont {O.~K.}\ \bibnamefont
  {Andersen}}, \bibinfo {author} {\bibfnamefont {O.}~\bibnamefont {Jepsen}}, \
  and\ \bibinfo {author} {\bibfnamefont {D.}~\bibnamefont {Gl{\"{o}}tzel}},\
  }in\ \href@noop {} {\emph {\bibinfo {booktitle} {Highlights of Condensed
  Matter Theory}}},\ \bibinfo {series and number} {International School of
  Physics `Enrico Fermi', Varenna, Italy,},\ \bibinfo {editor} {edited by\
  \bibinfo {editor} {\bibfnamefont {F.}~\bibnamefont {Bassani}}, \bibinfo
  {editor} {\bibfnamefont {F.}~\bibnamefont {Fumi}}, \ and\ \bibinfo {editor}
  {\bibfnamefont {M.~P.}\ \bibnamefont {Tosi}}}\ (\bibinfo  {publisher}
  {North-Holland},\ \bibinfo {address} {Amsterdam},\ \bibinfo {year} {1985})\
  pp.\ \bibinfo {pages} {59--176}\BibitemShut {NoStop}%
\bibitem [{\citenamefont {Harrison}(2004)}]{harrison:elementary2004}%
  \BibitemOpen
  \bibfield  {author} {\bibinfo {author} {\bibfnamefont {W.~A.}\ \bibnamefont
  {Harrison}},\ }\href@noop {} {\emph {\bibinfo {title} {Elementary Electronic
  Structure}}}\ (\bibinfo  {publisher} {World Scientific},\ \bibinfo {address}
  {Singapore},\ \bibinfo {year} {2004})\BibitemShut {NoStop}%
\bibitem [{\citenamefont {Xu}\ \emph {et~al.}(1992)\citenamefont {Xu},
  \citenamefont {Wang}, \citenamefont {Chan},\ and\ \citenamefont
  {Ho}}]{xu:jphys1992}%
  \BibitemOpen
  \bibfield  {author} {\bibinfo {author} {\bibfnamefont {C.~H.}\ \bibnamefont
  {Xu}}, \bibinfo {author} {\bibfnamefont {C.~Z.}\ \bibnamefont {Wang}},
  \bibinfo {author} {\bibfnamefont {C.~T.}\ \bibnamefont {Chan}}, \ and\
  \bibinfo {author} {\bibfnamefont {K.~M.}\ \bibnamefont {Ho}},\ }\href
  {\doibase 10.1088/0953-8984/4/28/006} {\bibfield  {journal} {\bibinfo
  {journal} {J. Phys.: Condens. Matter.}\ }\textbf {\bibinfo {volume} {4}},\
  \bibinfo {pages} {6047} (\bibinfo {year} {1992})}\BibitemShut {NoStop}%
\bibitem [{\citenamefont {Pereira}\ \emph {et~al.}(2009)\citenamefont
  {Pereira}, \citenamefont {Castro~Neto},\ and\ \citenamefont
  {Peres}}]{pereira:prb2009}%
  \BibitemOpen
  \bibfield  {author} {\bibinfo {author} {\bibfnamefont {V.~M.}\ \bibnamefont
  {Pereira}}, \bibinfo {author} {\bibfnamefont {A.~H.}\ \bibnamefont
  {Castro~Neto}}, \ and\ \bibinfo {author} {\bibfnamefont {N.~M.~R.}\
  \bibnamefont {Peres}},\ }\href {\doibase 10.1103/PhysRevB.80.045401}
  {\bibfield  {journal} {\bibinfo  {journal} {Phys. Rev. B}\ }\textbf {\bibinfo
  {volume} {80}},\ \bibinfo {pages} {045401} (\bibinfo {year}
  {2009})}\BibitemShut {NoStop}%
\bibitem [{\citenamefont {Schwarz}\ \emph {et~al.}(2002)\citenamefont
  {Schwarz}, \citenamefont {Blaha},\ and\ \citenamefont
  {Madsen}}]{Schwarz:ComPhysComm2002}%
  \BibitemOpen
  \bibfield  {author} {\bibinfo {author} {\bibfnamefont {K.}~\bibnamefont
  {Schwarz}}, \bibinfo {author} {\bibfnamefont {P.}~\bibnamefont {Blaha}}, \
  and\ \bibinfo {author} {\bibfnamefont {G.}~\bibnamefont {Madsen}},\ }\href
  {\doibase http://dx.doi.org/10.1016/S0010-4655(02)00206-0} {\bibfield
  {journal} {\bibinfo  {journal} {Computer Physics Communications}\ }\textbf
  {\bibinfo {volume} {147}},\ \bibinfo {pages} {71 } (\bibinfo {year}
  {2002})}\BibitemShut {NoStop}%
\bibitem [{\citenamefont {Suzuura}\ and\ \citenamefont
  {Ando}(2002)}]{suzuura:prb2002}%
  \BibitemOpen
  \bibfield  {author} {\bibinfo {author} {\bibfnamefont {H.}~\bibnamefont
  {Suzuura}}\ and\ \bibinfo {author} {\bibfnamefont {T.}~\bibnamefont {Ando}},\
  }\href {\doibase 10.1103/PhysRevB.65.235412} {\bibfield  {journal} {\bibinfo
  {journal} {Phys. Rev. B}\ }\textbf {\bibinfo {volume} {65}},\ \bibinfo
  {pages} {235412} (\bibinfo {year} {2002})}\BibitemShut {NoStop}%
\bibitem [{\citenamefont {Castro~Neto}\ and\ \citenamefont
  {Guinea}(2007)}]{castro-neto:prb2007}%
  \BibitemOpen
  \bibfield  {author} {\bibinfo {author} {\bibfnamefont {A.~H.}\ \bibnamefont
  {Castro~Neto}}\ and\ \bibinfo {author} {\bibfnamefont {F.}~\bibnamefont
  {Guinea}},\ }\href {\doibase 10.1103/PhysRevB.75.045404} {\bibfield
  {journal} {\bibinfo  {journal} {Phys. Rev. B}\ }\textbf {\bibinfo {volume}
  {75}},\ \bibinfo {pages} {045404} (\bibinfo {year} {2007})}\BibitemShut
  {NoStop}%
\bibitem [{\citenamefont {Cappelluti}\ and\ \citenamefont
  {Profeta}(2012)}]{cappelluti:prb2012}%
  \BibitemOpen
  \bibfield  {author} {\bibinfo {author} {\bibfnamefont {E.}~\bibnamefont
  {Cappelluti}}\ and\ \bibinfo {author} {\bibfnamefont {G.}~\bibnamefont
  {Profeta}},\ }\href {\doibase 10.1103/PhysRevB.85.205436} {\bibfield
  {journal} {\bibinfo  {journal} {Phys. Rev. B}\ }\textbf {\bibinfo {volume}
  {85}},\ \bibinfo {pages} {205436} (\bibinfo {year} {2012})}\BibitemShut
  {NoStop}%
\bibitem [{Note4()}]{Note4}%
  \BibitemOpen
  \bibinfo {note} {Alternatively a uniform distribution over $[-d,d]$ can be
  used.}\BibitemShut {Stop}%
\end{thebibliography}
\end{document}